\documentstyle[aps,epsfig,prd]{revtex} 

\def\baselinestretch{1.1}
\newcommand{\beq}[1]{
\begin{equation}\label{#1}}
\newcommand{\bea}[1]{
\begin{eqnarray}\label{#1}}
\def\be{\begin{equation}}
\def\ee{\end{equation}}
\def\bea{\begin{eqnarray}}
\def\eea{\end{eqnarray}}
\def\lsim{\mathrel{\rlap{\lower4pt\hbox{\hskip1pt$\sim$}}
    \raise1pt\hbox{$<$}}}         
\def\gsim{\mathrel{\rlap{\lower4pt\hbox{\hskip1pt$\sim$}}
    \raise1pt\hbox{$>$}}}
\def\dr{\not\!}

\begin{document}
\thispagestyle{empty}

\begin{flushright}
JLAB-THY-06-476 \\
\today  %
\end{flushright}

\vspace{1cm}

\begin{center}
{\Large \bf 
 Higher Fock State Contributions to the Generalized Parton Distribution
of Pion }
\\[1cm]
\end{center}
\begin{center}
{CHUENG-RYONG JI$^1$, YURIY MISHCHENKO$^{1,2}$, 
ANATOLY RADYUSHKIN$^{2,3}$\footnote{Also at Bogoliubov Laboratory of Theoretical 
Physics, JINR, Dubna, Russian Federation}
} \\[2mm]  
 {\em $^1$Department of Physics, Box 8202, North Carolina 
 State University, Raleigh, NC 
27695-8202} \\[2mm]
{\em $^2$Theory Center, Jefferson 
Lab\footnote{Notice: This manuscript has been authored by 
The Southeastern Universities Research Association, Inc. 
under Contract No. DE-AC05-84ER40150 with the U.S. 
Department of Energy. The United States Government 
retains and the publisher, by accepting the article for publication, 
acknowledges that the United States Government retains a non-exclusive, 
paid-up, irrevocable, world wide license to publish or reproduce 
the published form of this manuscript, or allow others to do so, 
for United States Government purposes.}, Newport News, VA 23606}\\[2mm]
{\em $^3$Physics Department, Old Dominion University, Norfolk, VA 23529
}
\end{center}
\vspace{1cm}

\begin{abstract} 
We discuss the higher Fock state ($q\bar q g$) contributions to the
nonzero value of the pion GPD at the crossover point $x=\zeta$ between the
DGLAP and ERBL regions. Using the phenomenological light-front constituent
quark model, we confirm that the higher Fock state contributions indeed give
a nonzero value of the GPD at the crossover point.
Iterating the light-front quark model wavefunction of the lowest $q\bar q$ Fock state
with the Bethe-Salpeter kernel corresponding to 
 the one-gluon-exchange, we include all possible time-ordered
$q\bar q g$ Fock state contributions and obtain the pion GPD satisfying 
necessary sum rules
and continuity conditions.

  \vspace{.5cm}
\noindent {PACS number(s): 13.40.Gp, 13.60.Fz, 14.40.Aq}
\vspace{.5cm}
\end{abstract}
\vskip .2in
\newpage

\section{Introduction}
The Compton scattering provides a unique tool
for studying hadronic structure. In distinction to 
 hadronic form factors, the Compton amplitude probes the hadrons
through a coupling of two electromagnetic currents, and thus 
provides a wealth of additional information. 
When the initial photon is highly virtual while  the
final one is real, one arrives at the kinematics of deeply virtual
Compton scattering (DVCS), the process \cite{Ji:1996ek,Radyushkin:1996nd}   which  
 was a subject of 
intensive theoretical analysis during the last decade. 
Within the perturbative  Quantum ChromoDynamics (QCD),
it was established  
\cite{Muller:1998fv,Ji:1996ek,Radyushkin:1996nd,Ji:1998xh,Collins:1998be}
 that the most important contribution to the amplitude of this exclusive 
process is given
by the convolution of a hard quark propagator and a nonperturbative function
describing long-distance dynamics which is now known as a ``generalized parton
distribution (GPD)'' \cite{Muller:1998fv,Ji:1996ek,Radyushkin:1996nd}
(see \cite{Goeke:2001tz,Diehl:2003ny,Belitsky:2005qn} for recent detailed reviews). 
The GPDs serve as a generalization of the ordinary (forward) parton
distributions and provide much more direct and sensitive information on
the light-front (LF) wavefunctions of the target hadron than the hadron form
factors. In particular, the momentum of the ``probed quark'' in GPDs is
 kept fixed at longitudinal
momentum fraction $x$, while for the form factor it is integrated out
(due to the nonlocal current operator
$\bar{\psi}(0)\gamma^\mu\psi(z)$ for the GPDs in contrast to
the local vertex $\bar{\psi}(0)\gamma^\mu\psi(0)$ for the form factor).
Within the GPD approach, form factors are treated as just the lowest moments of the GPDs.
Because of being generalized amplitudes, the GPDs always involve
the nonvalence contributions due to the presence  of the  asymmetry
between the longitudinal 
initial $P$ and final $P'$ hadron momenta characterized by the ``skewness''
 parameter $\zeta=(P-P')^+/P^+$.
The kinematic region where the longitudinal momentum fraction
$x$ of the probed quark is greater than the skewness parameter $\zeta$
({\it i.e.} $1>x>\zeta$) is called ``DGLAP region''  
(the evolution pattern for GPDs there is similar
to the DGLAP evolution \cite{Gribov:1972ri,Altarelli:1977zs,Dokshitzer:1977sg} 
of ordinary parton distributions) 
 while the remaining part  $0<x<\zeta$
of the longitudinal momenta  $0<x<\zeta$ is called 
``ERBL region'' (GPDs in this case have ERBL-type 
evolution \cite{Efremov:1978rn,Lepage:1979zb} 
 characteristic of meson distribution amplitudes). 
In the framework of LF dynamics,
the DGLAP and ERBL regions have also been denoted as the valence and
nonvalence regions, respectively, because the
parton-number-changing nonvalence Fock-state contributions cannot be
avoided for $0<x<\zeta$ while only the parton-number-conserving valence
Fock-state contributions are needed for $1>x>\zeta$. Thus, it has been a
great challenge to calculate the
nonvalence contributions to the GPDs in the framework of LF quantization.

Although many recent theoretical 
endeavors~\cite{Diehl:1998kh,Diehl:2000xz,Brodsky:2000xy,Burkardt:2000za,Tiburzi:2001je,Choi:2001fc,Choi:2002ic}
have been made in describing the GPDs in terms of LF wavefunctions, 
the results have not yet been satisfactory enough for
practical calculations. In Refs.~\cite{Diehl:2000xz} and \cite{Brodsky:2000xy}, the 
nonvalence
contributions to the GPDs have been rewritten in terms of LF
wavefunctions with different parton configurations. However, the
representation given in Refs.~\cite{Diehl:2000xz} and \cite{Brodsky:2000xy} requires to 
find all
the higher Fock-state wavefunctions while there has been relatively little
progress in computing the basic wavefunctions of hadrons from first
principles. In Refs.~\cite{Burkardt:2000za} and \cite{Tiburzi:2001je}, the GPDs were expressed in
terms of LF wavefunctions,  but only within toy models such as
the 't~Hooft model of $(1+1)$-dimensional QCD~\cite{Burkardt:2000za} and the scalar
Wick-Cutkosky model~\cite{Tiburzi:2001je}.
While these toy model analyses are helpful
to gain physical insight on the properties of the GPDs (especially,
the time reversal invariance, the continuity at the crossover
between the DGLAP and ERBL regions, and the sum rule constrained
by the electromagnetic form factor), the real $(3+1)$-dimensional QCD
motivates us to come up with a more realistic model for the application
to the analysis of GPDs.

In an effort toward this direction, one of us
(C. Ji)  presented  an effective
treatment of handling the nonvalence contributions to the GPDs of the pion
\cite{Choi:2001fc,Choi:2002ic} using the  LF constituent quark model (LFQM),
which has been phenomenologically quite successful in describing
the space-like form factors for the electromagnetic and radiative decays of
pseudoscalar and vector mesons~\cite{Choi:1997iq,Choi:1997qh,Kisslinger:2001gw} and the
time-like weak form factors for exclusive semileptonic and
rare decays of pseudoscalar mesons~\cite{Ji:2000fy,Choi:1999nu,Choi:2001hg}.
Our effective treatment of handling the nonvalence contributions is based
on the covariant Bethe-Salpeter (BS) approach formulated in the LF
quantization~\cite{Ji:2000fy} which we call LFBS approach. It  has been previously
applied to the exclusive semileptonic and rare decays of pseudoscalar
mesons~\cite{Choi:1999nu,Choi:2001hg} providing reasonable results compared to the data.
The projection of the four-dimensional two-body BS equations at the light-front
hypersurface has also been discussed in Ref.~\cite{FSCS} and the problem of
constructing gauge invariant current in terms of LF bound-state wavefunctions
has been handled in Ref.~\cite{KB}

In Ref.\cite{Choi:2001fc},  an artifact of discontinuity at $x=\zeta$
between valence and
nonvalence parts of the GPD had occurred but it was later cured in
the framework of the same LFBS approach by taking into account  both the 
vertices of the meson and of the gauge boson \cite{Choi:2002ic}. 
Due to the  consistent treatment of the 
vertices, the continuity of the GPD
at the crossover point $x=\zeta$ was secured. 
However, the value of the GPD  at the crossover point
vanished due to the
end-point behavior of the LF wavefunctions of
meson and gauge boson. In Ref.\cite{Choi:2002ic}, it was noted that
it may
be possible to avoid this effect 
of GPDs' vanishing at the crossover point  
by considering contributions of the 
higher Fock states
in the DVCS amplitude.
In this work, we examine if the GPD value at the crossover point is indeed
nonzero including the
higher Fock state corrections.
This investigation is particularly important because the recent measurement
of the single spin asymmetry (SSA) at HERMES \cite{Airapetian:2001yk} and
CLAS \cite{Stepanyan:2001sm} indicate  that the value of GPD at the crossover point
does not vanish for the proton. Although the 
internal structure of the pion is rather different
from that of  the proton, the issue 
of GPDs' vanishing or not vanishing 
 at the crossover point is common to all bound states.
 Thus, establishing 
the link between
the nonzero value of GPDs at the crossover point 
and the higher Fock-state contributions  is a  result 
applicable for both hadrons.

Working in the same LFBS framework, we will consider in addition
to the pion LF wavefunction the contribution from the higher Fock components of 
$q{\bar q}g$. We first follow the analysis of the higher 
Fock state contributions
presented in Ref.\cite{Tiburzi:2001je} and 
find that the analysis of Ref.\cite{Tiburzi:2001je} is 
correct only if one uses the exact solution
of the BS equation. 
Later, the authors of Ref.~\cite{Tiburzi:2001je} investigated the current matrix 
elements in the LFBS formalism discussing the replacement of non-wavefunction
vertices~\cite{TM1} and applying it to GPDs~\cite{TM2}. Using a simple model,
they obtained expressions for GPDs that are continuous and noted that the nonvanishing
of GPDs at the crossover point is tied to higher Fock components. Our work
develops the QCD application to the continuous and nonvanishing GPDs at the crossover
point. We present explicit numerical results in QCD.
For a model LF wavefunction we show that 
one shall retain all contributions including
the ones that ideally could be absorbed into the LF wavefunction
through the Bethe-Salpeter equation. This 
suggests a necessary improvement of the 
model wavefunctions by iterating them with the BS kernel.

This work is organized as follows. In Section \ref{sec2}, 
we go over briefly the essentials
of the LF kinematics of the DVCS. In Section \ref{sec3},
we review previous advances in LFBS effective treatment of the pion GPD.
Then, in Section \ref{sec4}, we extend LFBS approach to include
$q{\bar q}g$ Fock states and present the details of
our calculations and numerical results including the
application of the prescription given in Ref.\cite{Tiburzi:2001je}.
Concluding remarks follow in Section \ref{sec4}. 
The pole assignment for the Cauchy integration and the
comments on the organization of the numerical calculations are presented
in Appendices \ref{app1} and \ref{app2}, respectively.

\section{Light-Front Kinematics of the Deeply Virtual Compton Scattering}\label{sec2}

\begin{figure}[b]
\begin{center}
\epsfig{width=250pt,file=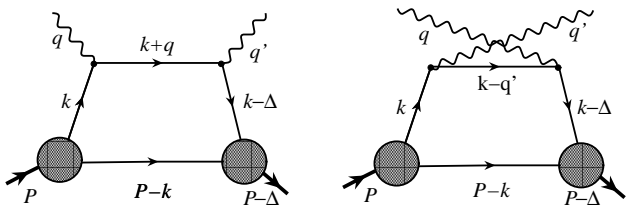}
\end{center}\vspace{0.3cm}
\caption{Handbag diagrams that give the dominant contribution  to Compton scattering
in the deeply virtual regime. The lower ``soft" part consists of a hadronic
matrix element which is parametrized by generalized
parton distribution functions.}
\label{handbag}
\end{figure}

We begin with the kinematics of the virtual Compton
scattering off the pion (see Fig.\ref{handbag})
\begin{equation}\label{VCS}
\gamma^*(q) + \pi(P) \to \gamma(q') + \pi(P').
\end{equation}

The initial (final) hadron state is characterized by the momentum
$P$ $(P')$, and the incoming  virtual (space-like)  
and outgoing real photons
are characterized by the 
momenta $q$ and $q'$, respectively. 
In this work, we use the LF-metric $V \cdot V=V^+V^- - {\bf V}_{\perp}^2$.
Defining the four momentum transfer $\Delta=P-P'$, one has
$$
P =\biggl (P^+, M^2/P^+, 0_\perp \biggr ) \ \ , \ \ 
P'= \biggl ((1-\zeta)P^+, (M^2+{\bf \Delta}^{2}_{\perp})/(1-\zeta)P^+, 
-{\bf \Delta}_\perp \biggr )
\ ,$$
and
$$\Delta = P - P' =(\zeta P^+, (\Delta^2+{\bf \Delta}^2_\perp)/\zeta P^+,
{\bf  \Delta}_\perp) \ , $$
where $M$ is the mass of the pion and $\zeta=\Delta^+/P^+$ is
the skewness parameter describing the longitudinal momentum asymmetry of  GPDs.
The squared momentum transfer is given by
\begin{equation}\label{del2}
t = \Delta^2 = 2 P\cdot\Delta=-\frac{\zeta^2 M^2 + {\bf \Delta}^2_\perp}{1-\zeta} \ .
\end{equation}
Since ${\bf \Delta}^2_\perp\geq 0$
at given $\zeta$, the value of $-t$ is constrained from below:
$-t>-t_{\min}=\zeta^2 M^2/(1-\zeta)$.

\noindent
As shown in Fig.\ref{handbag}, the parton emitted by the pion
has the momentum $k$, and the absorbed parton has the
momentum $k'=k-\Delta$.
Just like  in the case of space-like form factors, we may choose a frame
where the incident space-like photon has zero plus component, $q^+=0$, so that
\begin{equation}
\begin{array}{c}
q = \left (0,{ ({\bf q}_\perp + {\bf \Delta}_\perp)^2}/{\zeta P^+}
+ (\zeta M^2 +{\bf \Delta}^2_\perp)/{(1-\zeta) P^+}, {\bf q}_\perp \right ) \  \ , \ \ 
q' = \left ( \zeta P^+, {({\bf q}_\perp + {\bf \Delta}_\perp)^2}/{\zeta P^+} ,
{\bf q}_\perp +{\bf \Delta}_\perp \right ).
\end{array}
\end{equation}

\noindent
In DVCS, where $Q^2=-q^2 \gg M^2$, and $Q^2 \gg -t$ is large,
$\zeta$ plays the role of the
Bjorken variable, {\it i.e.} $Q^2/(2P\cdot q)=\zeta$. For a fixed
value of $-t$, the allowed range of $\zeta$ is given by
\begin{equation}\label{range}
0\leq\zeta\leq\frac{(-t)}{2M^2} \, \biggl(
\sqrt{1 + \frac{4M^2}{(-t)}}-1\biggr).
\end{equation}

\noindent
In the leading twist, ignoring interactions at the
quark-gauge-boson (photon in this case) vertex,
the dominant contribution   to the Compton scattering amplitude
in the deeply virtual region is given by 
\begin{equation}\label{Mmn}
M^{\mu\nu}=M^{\mu\nu}_s + M^{\mu\nu}_u \ , 
\end{equation}
{\it i.e.} the sum of the $s$-channel amplitude $M^{\mu\nu}_s$ and 
the $u$-channel amplitude $M^{\mu\nu}_u$ shown in Fig.\ref{handbag}.
The $s$-channel amplitude is given by
\begin{eqnarray}\label{Mmns}
M^{\mu\nu}_s&=& -iN_ce^2_q\int\frac{d^4k}{(2\pi)^4}
{\rm Tr} \, [\gamma_5(\not\!k +m)\gamma^\mu(\not\!k+\not\!q+m)\gamma^\nu
(\not\!k-\not\!\Delta+m)\gamma_5(-\not\!P+\not\!k+m)]
\nonumber\\
&\times&
\frac{H_{\rm cov}(k,P)H'_{\rm cov}(k-\Delta, P-\Delta)}
{[k^2-m^2+i\varepsilon][(k+q)^2-m^2+i\varepsilon]
[(k-\Delta)^2-m^2+i\varepsilon][(P-k)^2-m^2+i\varepsilon]},
\end{eqnarray}
where $N_c$ is the color factor and
$H_{\rm cov}(k,P)$ $[H'_{\rm cov}(k-\Delta,P-\Delta)]$ is the covariant
initial [final] state meson-quark vertex function that satisfies the
BS equation. As usual in the LFBS formalism,
we assume that the covariant vertex function $H_{\rm cov}(k)$
does not alter the $k^-$ pole structure in Eq.~(\ref{Mmns}).
The $u$-channel amplitude can be easily obtained by
$M^{\mu\nu}_u=M^{\mu\nu}_s(q\to-q')$.

\begin{figure}
\begin{center}
\epsfig{width=250pt,file=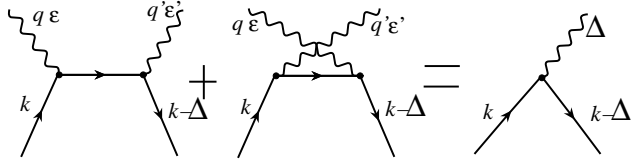}
\end{center}
\caption{Effective vertex}
\label{effDia}
\end{figure}

The effective gauge-boson vertex is depicted in Fig.\ref{effDia} as 
a reduction from the Compton scattering amplitude 
of the virtual photon given by
\be\label{redver1}
(-i e_q)^2 \left (\frac{\dr \epsilon i (\dr q + \dr k +m) \dr \epsilon '}
{(q+k)^2-m^2+i \varepsilon}+
\frac{\dr \epsilon ' i (-\dr q ' + \dr k +m) 
\dr \epsilon}{(k-q')^2-m^2+i \varepsilon}\right ) \ .
\ee
In the deep inelastic limit, we may neglect $\dr k$ along with
any 4-products not involving $q$ in the trace of the numerator 
(note that for circularly polarized photons
also $q\cdot \epsilon$ and $q \cdot \epsilon '$ can be neglected), {\it i.e.}
\be
q^2, q\cdot a \gg a\cdot b, m^2$, $ \epsilon \cdot q, \epsilon' \cdot q.
\ee
Then the amplitude (\ref{redver1}) can be rewritten as
\be\label{redver2}
 -i e_q^2 \left (\frac{-\dr q \dr \epsilon \dr \epsilon '}{(q+k)^2-m^2+i \varepsilon}+
\frac{\dr q'\dr \epsilon ' \dr \epsilon}{(k-q')^2-m^2+i \varepsilon} \right ) \, .
\ee
Furthermore, $\dr \epsilon \dr \epsilon '$ can be written as
$\epsilon \cdot \epsilon ' + i \sigma^{\mu\nu} \epsilon_\mu \epsilon'_\nu$,
where $\sigma^{\mu\nu}=\frac 1{2 i}[\gamma^{\mu},\gamma^{\nu}]$.
The axial term can be neglected because it vanishes 
after integration over ${\bf k}_\perp $. Thus,   for the reduced 
vertex we obtain 
\begin{eqnarray} 
 -i e_q^2  (-\epsilon \cdot \epsilon') \dr q \left(\frac{1}{(q+k)^2-m^2+i \varepsilon}-
\frac{1}{(k-q')^2-m^2+i \varepsilon} \right ) \sim \nonumber \\
 -i e_q^2 (-g^{\mu\nu}) \dr q \left (\frac{1}{(q+k)^2-m^2+i \varepsilon}-
\frac{1}{(k-q')^2-m^2+i \varepsilon}\right ) \, .\label{redver3}
\end{eqnarray}
The denominators can be further simplified by
\be
 \left ( \frac{1}{(q+k)^2-m^2+i \varepsilon}-
\frac{1}{(k-q')^2-m^2+i \varepsilon} \right ) \sim 
\frac{1}{P^+ q^-} \left (\frac 1{x-\zeta+i\varepsilon} +
 \frac{1}{x-i\varepsilon} \right ) \ .
\ee
Adding $s$- and $u$-channel amplitudes,
we obtain the Compton scattering amplitude in DVCS limit as follows
\begin{eqnarray}\label{CDVCS}
M^{IJ}= \epsilon^I_\mu\epsilon^{*J}_\nu M^{\mu\nu}
&=&-\frac{iN_c}{2P^+}e^2_q\int\frac{d^4k}{(2\pi)^4}
\biggl(\frac{1}{x-\zeta+i\varepsilon}+
\frac{1}{x-i\varepsilon}\biggr)
H_{\rm cov}(k,P)H'_{\rm cov}(k-\Delta, P-\Delta)
\nonumber\\
&\times&
\frac{
{\rm Tr} \, [\gamma_5(\not\!k +m)
\not\!\epsilon^I\gamma^+\not\!\epsilon^{*J}
(\not\!k-\not\!\Delta+m)\gamma_5(-\not\!P+\not\!k+m)] }
{[k^2-m^2+i\varepsilon] [(k-\Delta)^2-m^2+i\varepsilon]
[(P-k)^2-m^2+i\varepsilon]}.
\end{eqnarray}
For circularly polarized ($\epsilon^+=0$) initial and final
photons\footnote{As discussed in~\cite{Brodsky:2000xy}, for a longitudinally
polarized initial photon, the Compton amplitude is of order $1/Q$
and thus vanishes in the limit $Q^2\to\infty$.}($I,J$ are $\uparrow$
or $\downarrow$),
we obtain from Eq.~(\ref{CDVCS})
\begin{equation}\label{pol}
\not\!\epsilon^I\gamma^+\not\!\epsilon^{*J}
= (\epsilon^I_\perp\cdot\epsilon^{*J}_\perp)\gamma^+
+ i(\epsilon^I_\perp\times\epsilon^{*J}_\perp)_3\gamma^+\gamma_5.
\end{equation}
Equation (\ref{pol}) reduces to $\gamma^+(1\pm\gamma_5)$ for
the parallel helicities ({\it i.e.} $+$ for $\uparrow\uparrow$ and
$-$ for $\downarrow\downarrow$) and zero otherwise.
Since the axial current $\gamma^+\gamma_5$ does not contribute
to the integral, {\it i.e.} $\sim({\bf k}_\perp\times{\bf \Delta}_\perp)$ after
the trace calculation, this term can be dropped.

Thus, the DVCS amplitude ({\it i.e.} photon helicity amplitude)
can be rewritten as the factorized form of hard and soft amplitudes
\begin{equation}\label{CDVCS2}
M^{\uparrow\uparrow}(P,q,P') = M^{\downarrow\downarrow}(P,q,P')
=-e^2_q\int\; dx \biggl(\frac{1}{x-\zeta+i\varepsilon}+
\frac{1}{x-i\varepsilon}\biggr){\cal F}_\pi(\zeta,x,t),
\end{equation}
where
\begin{eqnarray}\label{HPI}
{\cal F}_\pi(\zeta,x, t)=
&=&\frac{iN_c}{2}\int\frac{dk^-d^2{\bf k}_\perp}{2(2\pi)^4}
H_{\rm cov}(k,P)H'_{\rm cov}(k-\Delta, P-\Delta)
\nonumber\\
&\times&
\frac{
{\rm Tr} \, [\gamma_5(\not\!k +m)\gamma^+
(\not\!k-\not\!\Delta+m)\gamma_5(-\not\!P+\not\!k+m)] }
{[k^2-m^2+i\varepsilon] [(k-\Delta)^2-m^2+i\varepsilon]
[(P-k)^2-m^2+i\varepsilon]}.
\end{eqnarray}
The function ${\cal F}_\pi(\zeta,x,t)$ is the
``generalized parton distribution'' and it manifests characteristics of
the ordinary(forward) quark distribution in the limits $\zeta\to 0$
and $t\to 0$. On the other hand,  the first moment of
${\cal F}_\pi(\zeta,x,t)$ 
is related to the form factor by the
following sum rule \cite{Ji:1996ek,Radyushkin:1996nd}:
\begin{equation}\label{sum}
\int^1_{0} {dx}\, 
{\cal F}_{\pi}(\zeta, x, t)
=(1-{\zeta}/{2}) \, F_\pi(t).
\end{equation}
In general, the polynomiality conditions for the moments of
the GPDs\cite{Ji:1997gm,Radyushkin:1998bz} defined by
\begin{equation}\label{sum-mom}
\int^1_{0}  dx \,  x^{n-1} {\cal F}_{\pi}(\zeta, x, t)=(1-{\zeta}/{2})\, F_n(\zeta,t)
\end{equation}
require that the highest power of $\zeta$ in the polynomial expression of
$F_n(\zeta,t)$ should not be larger than $n$. These polynomiality
conditions are fundamental properties of the GPDs which follow from the
Lorentz invariance. The positivity property of the GPDs,
following from positivity of the density
matrix \cite{Radyushkin:1998es}, has been discussed
analyzing the double distributions (DD) \cite{Tiburzi:2002kr}. While
such conditions as sum rules and
polynomiality are more easily satisfied by the DD-based models for the GPDs,
it has been noted\cite{Tiburzi:2002kr} that the positivity constraints are more transparent 
in the framework based on the LF wavefunction.

An important feature of the DVCS amplitude given by Eq.~(\ref{CDVCS2})
is that it depends only on the skewness parameter $\zeta=Q^2/(2P\cdot q)$
for large $Q^2$ and fixed $|t|(\leq Q^2)$. 
This  property of the exclusive DVCS  amplitude is similar 
to the Bjorken scaling  for the 
structure functions 
of the inclusive deep inelastic scattering (DIS) reaction.
Moreover, for DVCS, 
 the skewness parameter   $\zeta$ 
coincides with the Bjorken variable $x_{Bj}$. 
Note also that,  according to 
 Eq.~(\ref{CDVCS2}), the size  of the 
 imaginary part of the DVCS
amplitude is proportional to   
 ${\cal F}_\pi(\zeta,\zeta,t)$, {\it i.e.}, to the value 
  of the GPD at the crossover point 
 $x=\zeta$.    
In its turn, the imaginary part 
of the DVCS amplitude determines  the magnitude of 
the single spin asymmetry (SSA) \cite{Ji:1996ek}
(see also \cite{Kivel:2000fg,Freund:2001hd})  that can be measured through  the
scattering of  longitudinally polarized electrons  on an unpolarized target. 
The  SSA measurements  for the
proton target have been reported by HERMES \cite{Airapetian:2001yk} 
and CLAS \cite{Stepanyan:2001sm} 
collaborations, with the magnitude of the SSA  
definetely inconsistent with zero.   

\section{Light Front Bethe-Salpeter approach for the 
generalized parton distributions.} \label{sec3}

\begin{figure}
\centering
\begin{minipage}[c]{0.35\hsize}
\epsfig{width=\hsize,file=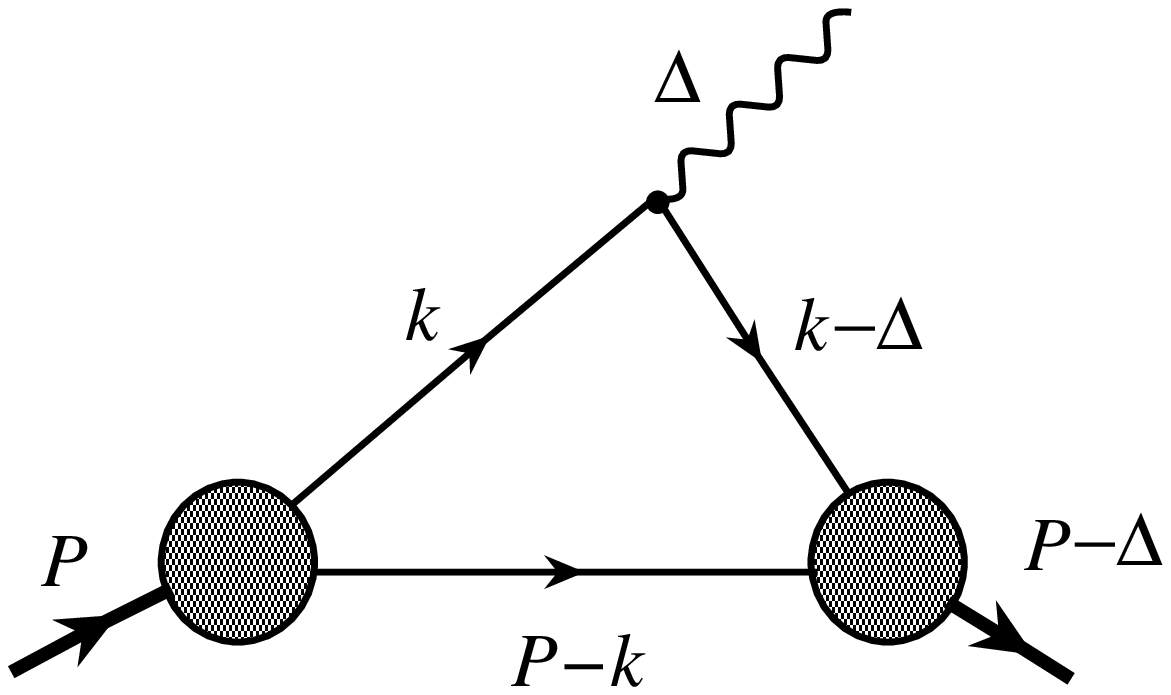}
\end{minipage}
\hspace*{0.25cm}
\begin{minipage}[c]{0.35\hsize}
\epsfig{width=\hsize,file=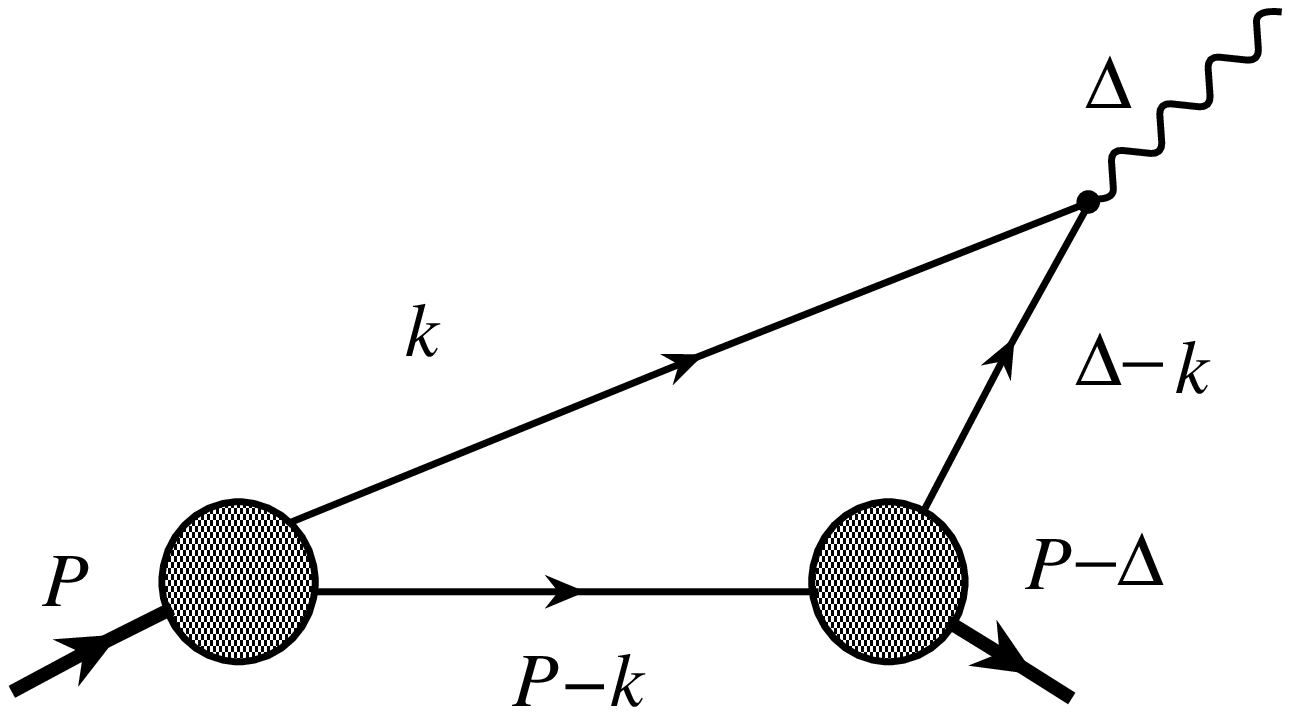}
\end{minipage}\vspace{0.3cm}
\caption{Covariant amplitude with reduced photon vertex for pion GPD (left)
and its 
non-valence $x<\zeta$ part (right).
}\label{S0}
\end{figure}

Let us now consider an explicit form of the GPD defined by Eq.~(\ref{HPI}) in
the framework of Light-Front  dynamics. 
Corresponding to the $x<\zeta$ and $x>\zeta$ regions of the covariant
amplitude (see Fig.\ref{S0}),
the Cauchy integration over $k^-$ in Eq.~(\ref{HPI}) has nonzero
contribution coming either from the residue of the pole at 
$(P-k)^2=m^2$ in the valence region $x>\zeta$ or
from the pole at $k^2=m^2$ in the non-valence region $x<\zeta$.

These two cases correspond to two different time-ordered LF diagrams
which we will consider separately.
In the region of $\zeta<x$ the residue is taken at the pole in upper
half-plane $k^-$ relevant to $(P-k)^2=m^2$, {\it i.e.} 
$k^-=P^- -(P- k)^-_{\rm on}+i\epsilon=P^- - [{\bf k}^2_\perp + m^2]/(P-k)^++i\epsilon$. 
Performing  the Cauchy integration over $k^-$ in this region we find 

\begin{eqnarray}
k^2-m^2=k^+k^--{\bf k}_\perp ^2-m^2\rightarrow
k^+(P^- - k^-_{\rm on}-(P-k)^-_{\rm on}) = x 
\left (M^2 - \frac{{\bf k}^2_\perp  + m^2}{x(1-x)} \right  ) \ , 
\nonumber \\
(k-\Delta)^2-m^2\rightarrow(k-\Delta)^+(P^- - \Delta^- - 
(P-k)^-_{\rm on} - (k-\Delta)^-_{\rm on}) =
x^{\prime}
\left (M^2 - \frac{k^{\prime 2}_\perp + m^2}{x^\prime (1-x^\prime)} \right )  \ ,
\end{eqnarray}
where
\begin{equation}\label{xpkp}
x'= \frac{x-\zeta}{1-\zeta} \ \ \text{ and } \ \ 
{\bf k'}_\perp={\bf k}_\perp -{\bf \Delta}_\perp+x' {\bf \Delta}_\perp
\end{equation}
are the ``internal'' LF-momenta of the quark absorbed into the final pion state
($x'$ evidently has the meaning of the plus-momentum of the 
absorbed quark measured in units of the final hadron plus-momentum).  
Thus, the Cauchy integration 
in Eq.~(\ref{HPI}) over $k^-$ gives
\be\label{jv}
{\cal F}^{\rm val}_\pi(\zeta,x,t) =
\frac{N_c}{2 P^+}
\frac{1}{16\pi^3}\frac{1}{(1-x) x x'}
\int d^2{\bf k}_\perp \, 
\chi_{(2\to 2)}(x,{\bf k}_\perp) \, S^+_{\rm val} \, 
\chi'_{(2\to 2)}(x',{\bf k}'_{\perp}),
\ee 
where 
\bea\label{wfI}
\chi_{(2\to 2)}&=&\frac{h_{LF}}{M^2 -M^2_0} \ \ , \ \ 
M^2_0=\frac{{\bf k}^2_\perp + m^2}{1-x}
+ \frac{{\bf k}^2_\perp + m^2}{x},\nonumber\\
\chi'_{(2\to 2)}&=&\frac{h'_{LF}}{M^2 -M'^2_0} \ \ , \ \ 
M'^2_0=\frac{{\bf k'}^2_\perp + m^2}{1-x'}
+ \frac{{\bf k'}^2_\perp + m^2}{x'},\nonumber 
\eea
and $S^+_{\rm val}$ is the trace in Eq.~(\ref{HPI}) evaluated at the point
$(P-k)^2=m^2$;~{\it i.e.}
\begin{equation}\label{SV}
S^+_{\rm val}=\frac{4 P^+}{1-x'} \, ({\bf k}_\perp\cdot{\bf k'}_\perp + m^2).
\end{equation}
The light-front vertex functions $h_{LF}$ and $h'_{LF}$ can be related to the 
covariant $H_{\rm cov}$ and $H'_{\rm cov}$ vertex functions and 
then to LF wavefunctions, e.g.,
via instantaneous approximation following \cite{Tiburzi:2001je}. 
Here, following \cite{Choi:2002ic}, we identify
the LF wavefunction in $\chi_{(2\to 2)}$ as 
\begin{equation}\label{LFvertex}
\chi_{(2\to 2)}(x,{\bf k}_\perp)=
\sqrt{\frac{8\pi^3}{N_c}}
\sqrt{\frac{\partial k_z}{\partial x}} \, 
\frac{[x(1-x)]^{1/2}}{M_0} \, \phi(x,{\bf k}_\perp),
\end{equation}
where the Jacobian of the  transformation 
${\bf k}=(k_z,{\bf k}_\perp)\to
(x,{\bf k}_\perp)$ is obtained using $\partial k_z/\partial x=M_0/[4x(1-x)]$
with $M_0=\sqrt{\frac{{\bf k}_\perp^2+m^2}{x(1-x)}}$,
and the radial wavefunction is given by 
\begin{equation}\label{radial}
\phi(x,{\bf k}_\perp^2)=\chi \, 
\sqrt{\frac{1}{\pi^{3/2}\beta^3}}
\exp\left\{ \frac{M^2-({\bf k}^2_\perp+m^2)/(x(1-x))}{8\beta^2}\right\}.
\end{equation}   
The normalization factor $\chi=1$ is introduced for future convenience
and is fixed by the normalization of the pion form-factor $F_\pi(t=0)=1$.
After some simplifications, Eq.~(\ref{LFvertex}) can be written as
\begin{equation}\label{LFvertex1}
\chi_{(2\to 2)}(x,{\bf k}_\perp)=
\sqrt{\frac{2\pi^3}{N_c}}
\frac{1}{\sqrt{M_0}}\phi(x,{\bf k}_\perp).
\end{equation}

\noindent
Substituting Eq.~(\ref{LFvertex}) into 
Eq.~(\ref{jv}), we obtain the valence
part of the pion GPD:
\bea\label{light-frontFv}
{\cal F}^{\rm val}_\pi(\zeta,x,t)&=&
\int d^2{\bf k}_\perp
\sqrt{\frac{\partial k'_z}{\partial x'}}
\sqrt{\frac{\partial k_z}{\partial x}}
\phi(x',{\bf k'}_\perp)
\phi(x,{\bf k}_\perp)\frac{({\bf k}_\perp\cdot{\bf k'}_\perp + m^2)}
{\sqrt{{\bf k}^2_\perp + m^2}\sqrt{{\bf k'}^2_\perp + m^2}}. 
\eea

In the region of $x<\zeta$, the residue shall be taken at the pole
$k^2=m^2$, {\it i.e.}
$k^-=k^-_{\rm on}=[{\bf k}^2_\perp + m^2]/k^+-i\epsilon$, which is located 
in the lower half of the complex-$k^-$ plane. 
One can show that in this case
\begin{eqnarray}
(P-k)^2-m^2\rightarrow(1-x)\left (M^2 - \frac{{\bf k}^2_\perp  + m^2}{x(1-x)}\right ), 
\nonumber \\
(k-\Delta)^2-m^2\rightarrow(1-x'')
\left (\Delta^2-\frac{{\bf k}_\perp^{\prime\prime 2}+m^2}{x''(1-x'')} \right ),
\label{NVPI}
\end{eqnarray}
where
\begin{equation}
x''=x/\zeta $ \ \ and  \ \ $ {\bf k}''_\perp = {\bf k}_\perp + x'' {\bf \Delta}_\perp
\end{equation}
are the ``internal'' momenta of 
the quark annihilating into the photon. The second line in
Eq.~(\ref{NVPI}) is related to a ``gauge-boson-wavefunction''
 $\chi ^g$ \cite{Choi:2001fc}.

Then the non-valence part of GPD after the Cauchy integration in Eq.~(\ref{HPI})
can be written as 
\be \label{jnv}
{\cal F}^{\rm nonval}_\pi(\zeta,x,t)=
\frac{N_c}{2P^+}
\frac{1}{16\pi^3}\frac{1}{x(1-x)(1-x'')} \int d^2{\bf k}_\perp \, 
\chi_{(2\to 2)}(x,{\bf k}_\perp)S^+_{\rm nonval} \, \chi^g(x,{\bf k''}_\perp) \, h'_{LF},
\ee
where $\chi_{2\rightarrow 2}$ and $\chi^g$ are the pion and gauge-boson
LF wavefunctions and the final state LF vertex function $h'_{LF}$ is now kept 
explicitly.
The trace term at $k^2=m^2$ ($S^+_{\rm nonval}$) is given by 
\begin{equation}\label{snv}
S^+_{\rm nonval}= S^{+}_{\rm val} + \frac{4P^+}{1-x'}x(1-x)x'(M^2-M^2_0).
\end{equation}
\begin{figure}
\begin{center}
\epsfig{width=250pt,file=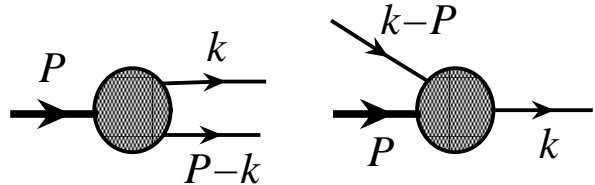}
\end{center}\vspace{0.3cm}
\caption{The difference in valence and non-valence LF vertex functions
$h^{\prime val}_{LF}$ and $h^{\prime nv}_{LF}$.}
\label{valnon}
\end{figure}
Unlike in the valence part of GPD, the final state vertex
$h'_{LF}$ cannot be simply reduced to a LF wavefunction
since it does not describe annihilation/creation of quark-antiquark pair
out of the pion state. It corresponds to  
the decay of a quark into a quark and a pion state, 
or, equivalently the annihilation of a pion and a quark into a quark (see Fig.\ref{valnon}).
However, it is worth noting that $h'_{LF}$ for valence and non-valence
contribution essentially are the special case of the same object, 
namely, the covariant vertex function $H(k,P)$,  in 
different kinematic regions. In particular, they 
obey the same BS equation and, in principle, can be
deduced from the usual LF wavefunction by analytic continuation, as
pointed out in Refs.\cite{Choi:2001fc,Choi:2002ic}.

\begin{figure}
\begin{center}
\begin{minipage}[c]{0.35\hsize}
\epsfig{width=\hsize,file=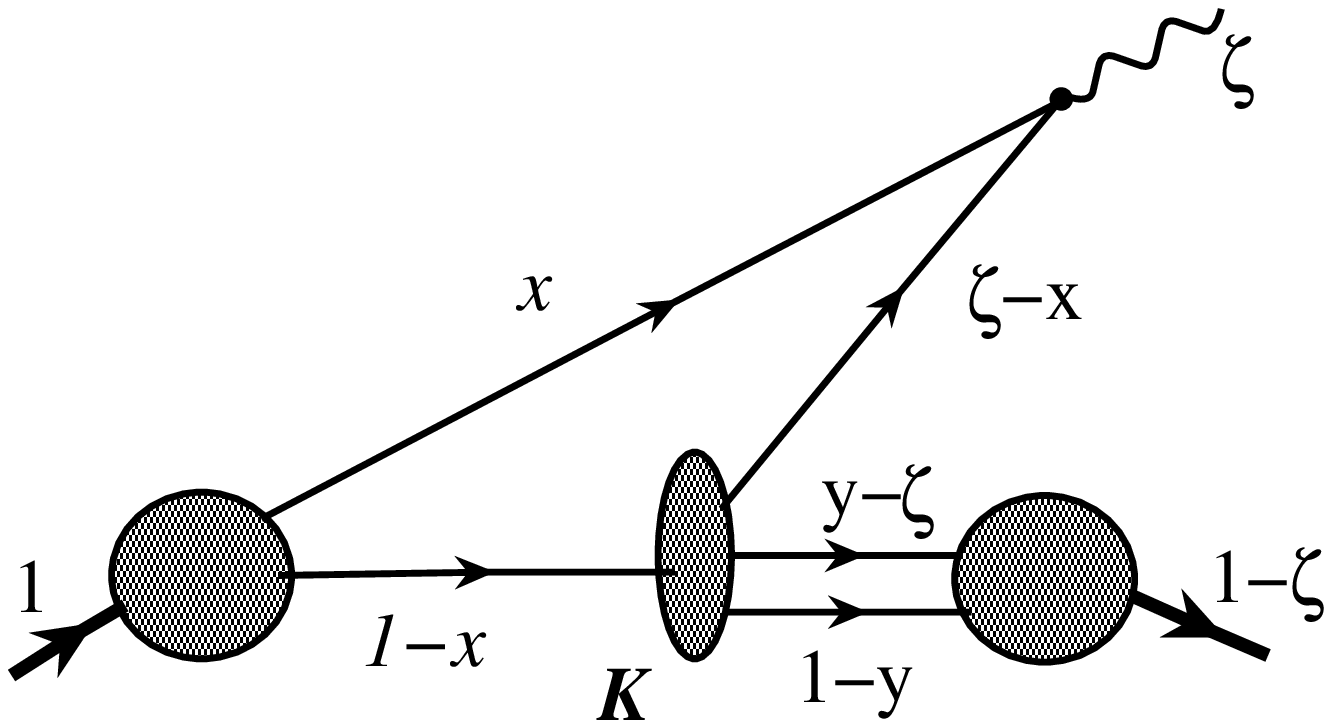}
\end{minipage}
\begin{minipage}[c]{0.35\hsize}
\epsfig{width=\hsize,file=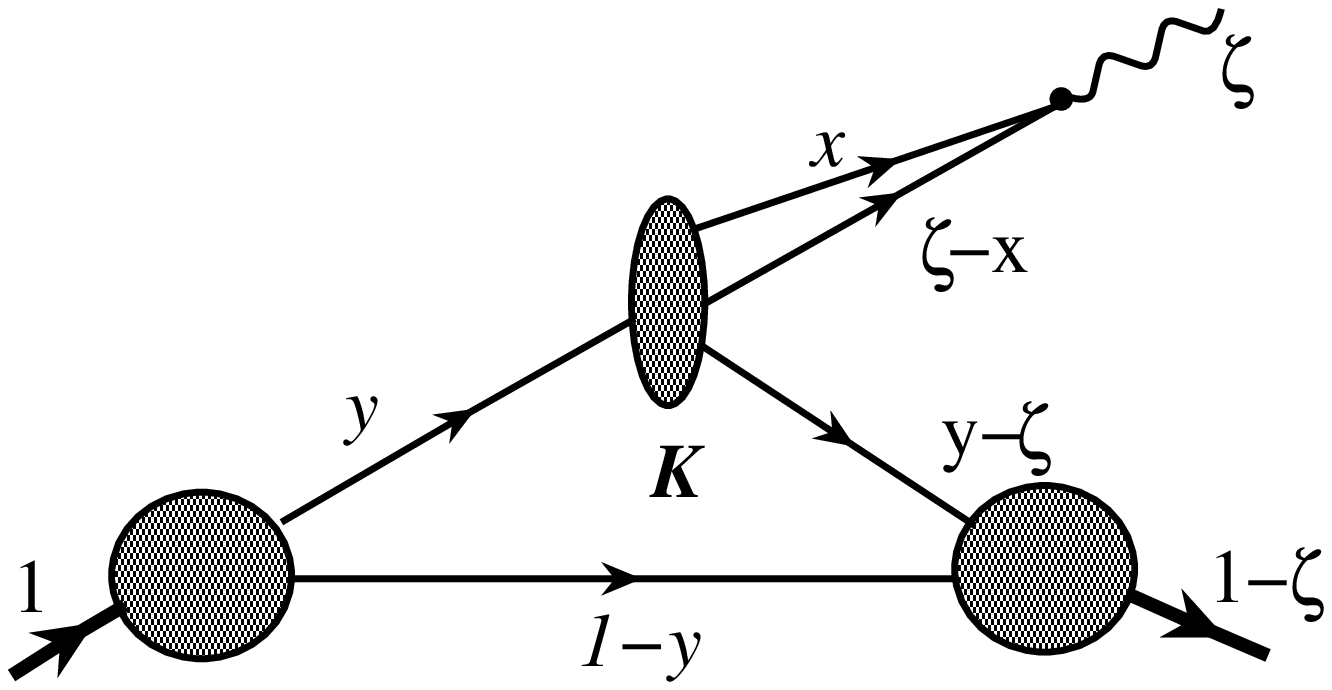}
\end{minipage}
\end{center}\vspace{0.3cm}
\caption{Non-valence part of GPD interpretation in LF CQM.}
\label{highFockb}
\end{figure}

Specifically, following Ref.\cite{Choi:2001fc},  the 
$\chi$-functions in LF CQM are given by the solutions of the BS 
equation \cite{Ji:2000fy,Brodsky:1984vp,Sales:1999ec}
\bea\label{SDtype1}
&&(M^2-M^2_0)\chi(x_i,{\bf k}_{i\perp})
\nonumber\\
&&=\int [dy][d^2{\bf l}_\perp] \, 
{\cal K}(x_i,{\bf k}_{i\perp}; y_j,{\bf l}_{j\perp}) \, 
\chi(y_j,{\bf l}_{j\perp}),
\eea
where ${\cal K}$ is the full BS kernel  (which in principle includes all the 
higher Fock-state contributions) and
\mbox{$M^2_0=(m^2+{\bf k}^2_{\perp})/(1-x) - (m^2 + {\bf k}^2_{\perp})/x$.} 
Both valence and nonvalence BS amplitudes 
satisfy Eq.~(\ref{SDtype1}).
For the usual BS amplitude, referred as the valence wavefunction, 
$x$ is greater than $\zeta$, while for the nonvalence BS amplitude $x$ is less than $\zeta$. We 
use the notation for these two solutions
\bea\label{BSamp}
   \chi_{(2\to 2)}& =& \chi^{\rm val}, \nonumber\\
 \chi_{(1\to 3)}& = & \chi^{\rm nonval}.
\eea
\begin{figure}
\begin{center}
\epsfig{width=250pt,file=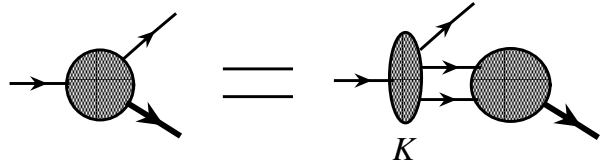}
\end{center}\vspace{0.3cm}
\caption{LF interpretation of the non-valence vertex ($1\rightarrow 3$ blob).}
\label{highFock}
\end{figure}
This notation is motivated by the relationship to the Fock state 
picture, in which  the parton number before and
after the kernel is interpreted for the nonvalence vertex as changing from 1 to 3.
According to \cite{Choi:2001fc,Choi:2002ic}, 
the nonvalence BS amplitude shall be expressed via the
valence BS amplitude and the full BS kernel in the relevant kinematic domain
as illustrated in Fig.\ref{highFock}:
\bea\label{SDtype}
&&(M^2-M^2_0)\chi_{(1\to 3)}(x_i,{\bf k}_{i\perp})
\nonumber\\
&&=\int [dy][d^2{\bf l}_\perp]
{\cal K}(x_i,{\bf k}_{i\perp}; y_j,{\bf l}_{j\perp})
\chi_{(2\to 2)}(y_j,{\bf l}_{j\perp}).
\eea  
In a sense, the BS kernel in this region can be viewed as
the sum of all processes resulting in creation of
a quark-antiquark pair from the initial single quark state.
With this in mind, we write the non-valence part of GPD 
(see left diagram in Fig.\ref{highFockb}) as 
\bea\label{jnv1}
{\cal F}^{\rm nonval}_\pi(\zeta,x,t)&=&\frac{N_c}{2P^+}
\frac{1}{16\pi^3}\frac{1}{x(1-x)(1-x'')}\int d^2{\bf k}_\perp
\chi_{(2\to 2)}(x,{\bf k}_\perp)S^+_{\rm nonval}\chi^g(x,{\bf k''}_\perp) 
\nonumber\\
& \times & \int\frac{dy}{y(1-y)}\int d^2{\bf l}_\perp
{\cal K}(x,{\bf k}_\perp; y,{\bf l}_\perp)\chi_{(2\to 2)}(y,{\bf l}_\perp).
\nonumber\\
\eea

It may be also noted that one should as well consider in this case the process
in which quark-antiquark pair is created not
by the spectator quark, but by the interacting quark (see right diagram in Fig.\ref{highFockb}).
In Ref.\cite{Choi:2002ic}, it was shown that this contribution
can be reduced to the form of Eq.~(\ref{jnv1}) with a simple
modification of the kernel
\bea\label{ktilde}
{\tilde{\cal K}}(x',{\bf k'}_\perp; y',{\bf l'}_\perp)
&\equiv& {\cal K}(x',{\bf k'}_\perp; y',{\bf l'}_\perp)
\biggl[
1-\frac{S^+_{\rm nonval}(y,{\bf l}_\perp){\tilde\chi}^{\rm i}_b(y,{\bf l}_\perp)}
{S^+_{\rm nonval}(x,{\bf k}_\perp){\tilde\chi}^{\rm i}_a(x,{\bf k}_\perp)}
\biggr],
\eea
where ${\tilde\chi}^{\rm i}_{a,b}$ denotes kinematic prefactors like
$x(1-x)(1-x'')$ etc. in the expressions corresponding to the diagrams
shown in
Fig.\ref{highFockb}.

\begin{figure}
\begin{center}
\epsfig{width=250pt,file=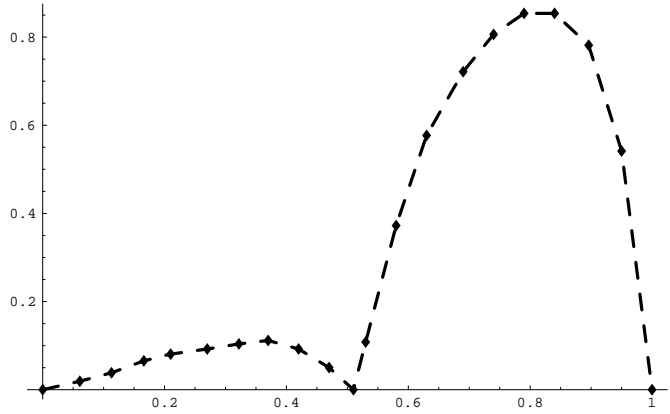}
\end{center}\vspace{0.3cm}
\caption{The pion GPD at $\zeta=0.5$, $t=-0.5$\,GeV$^2$ computed using LFBS
approach.}
\label{GPD0}
\end{figure}

Thus, we obtained the amplitude corresponding
to the nonvalence contribution given by Eq.~(\ref{jnv}) in terms of
ordinary light-front wavefunctions of hadron ($\chi_{(2\to 2)}$) and 
gauge-boson ($\chi^g$). 
This method, however, requires the knowledge of
the full BS kernel ${\cal K}(x,{\bf k}_\perp; y,{\bf l}_\perp)$ 
which is in general
dependent on the momenta connecting
the one-body to three-body sectors, as depicted in Fig.\ref{highFock}.
While the relevant operator ${\cal K}$
in general involves all momenta 
$(x,{\bf k}_\perp; y,{\bf l}_\perp)$, the integral of ${\cal K}$
over $y$ and ${\bf l}_\perp$ in Eq.~(\ref{jnv1}),
\bea\label{New_Gpi}
G_\pi&\equiv&
\int^1_0\frac{dy'}{y'(1-y')}\int d^2{\bf l}_\perp
{\tilde{\cal K}}(x,{\bf k}_\perp;
y',{\bf l'}_\perp) \chi_{(2\to2)}(y',{\bf l'}_\perp),
\eea
depends only on $x$ and ${\bf k}_\perp$. 
In this work, we approximate $G_\pi$ as a constant (mean value). This approximation
has been previously tested in the analyses of exclusive semileptonic decay 
processes \cite{Choi:2001fc,Choi:2002ic,Ji:2000fy}
and proved to be a good approximation at least in small momentum transfer 
region. 
The pion GPD, calculated in this way, 
looks like in Fig.\ref{GPD0}.

Using the ``effective'' gauge-boson wavefunction
in the non-valence region in the form 
\be\label{Gradial}
\phi^g({\bf k_{\perp}''}^2)=
\sqrt{\frac{1}{\pi^{3/2}\beta^3}}
\exp \left [- \frac{{\bf k}^{\prime\prime 2}_\perp+m^2}{8\beta^2x''(1-x'')}
\right  ]\exp \left (\frac{\Delta^2}{8\beta^2} \right ),
\ee
one can ensure the continuity between the valence and the nonvalence parts of the GPD
by suppressing the non-valence contribution at $x'=x''=0$ \cite{Choi:2002ic}. 
This cures  the  discontinuity found previously 
in Ref.\cite{Choi:2001fc} whose  origin
can be traced down to
the difference in the treatment of the valence and non-valence LF vertex
functions. More specifically, 
the discontinuity in Ref.\cite{Choi:2001fc} originated from the
crude approximation for the LF nonvalence vertex $h'^{\rm nonval}_{LF}$. As discussed in the 
general form above, the nonvalence vertex $h'^{\rm nonval}_{LF}$ is an analytic
continuation of $h'^{\rm val}_{LF}$ into nonvalence kinematic domain so that it should
be continuously and smoothly connected with the latter
at the boundary between the valence and the nonvalence kinematic regions. 
We experimented with different forms of such extensions for the non-valence vertex, 
satisfying condition of continuity at $x'=0$. One such possibility,
e.g., could be 
\be
h'^{\rm nonval}_{LF}\sim G_\pi \exp \left [\frac{{\bf k}_\perp ^2 +m^2}{8x\beta^2} \right ],
\ee
which   has  structure similar to that of the  LF wavefunction, 
goes to a constant when $x\ll -\frac{m^2}{\beta^2}$,
and vanishes at $x\rightarrow-0$, thus  guaranteeing the continuity with the valence 
LF wavefunction $h_{LF}^{\rm val}$.
However, the resultant behavior of the GPD in the non-valence
region depends significantly on the detailed form of the model assumed
for the extension of LF wavefunction into the 
non-valence region.
On the other hand,
while the introduction of
the model for the nonvalence vertex satisfying these requirements is
nontrivial,  the inclusion of the  virtual processes at the gauge-boson vertex
using Eq.~(\ref{Gradial}) removes the discontinuity at $x=\zeta$, thus alleviating 
the above problem at the lowest approximation of LFBS approach.
Such a treatment ensures the vanishing of the nonvalence contribution at the crossover
and also removes the infrared singularity in the amplitude.
It was further
suggested \cite{Choi:2002ic} that the value of the GPD at the crossover 
point needs not be zero because
the higher Fock states in LFBS approach 
may introduce a nonzero contribution at this point.
In the following section we will consider this suggestion in more detail and
show that, after inclusion  of the $q{\bar q}g$ contribution into the DVCS amplitude,
this value indeed is not zero.

\section{Effect of the higher Fock states in the LFBS approach for
the pion GPD.}
\label{sec4}

\begin{figure}
\begin{center}
\epsfig{width=150pt,file=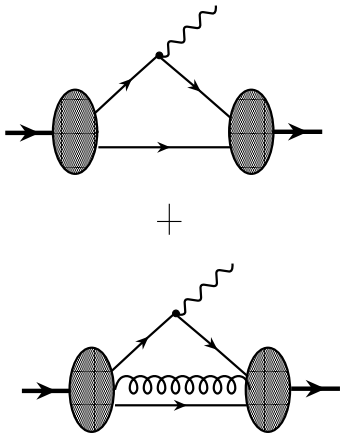}
\end{center}\vspace{0.3cm}
\caption{2-body and 3-body contributions to pion GPD in LF.}
\label{3states}
\end{figure}

Our primary goal in this section is to analyze how GPDs are affected by   
the higher
LF Fock states, especially at or near the crossover point $x=\zeta$. 
While the LFBS treatment in
Section \ref{sec3} can be considered as computing
GPD with 2-body Fock state contributions (upper diagram in Fig.\ref{3states}), 
here we are
interested in the effect of the 3-body Fock states as
depicted in the lower diagram in Fig.\ref{3states}. 
We concentrate specifically on the effect
of the Fock states that include a gluon in addition
to the constituent quarks in the DVCS amplitude.
As suggested in Ref.\cite{Choi:2002ic}, such contributions may
result in  a nonzero value at the crossover point of the GPD.

\begin{figure}
\begin{center}
\epsfig{width=250pt,file=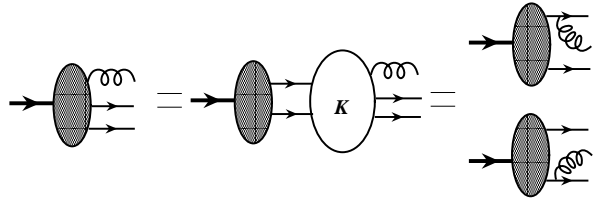}
\end{center}\vspace{0.3cm}
\caption{Relation between 2-body and 3-body Fock states (left) and
an approximate relation for the kernel $K$ (right).}
\label{23body}
\end{figure}

\begin{figure}
\centering
\begin{minipage}[c]{0.3\hsize}
\epsfig{width=\hsize,file=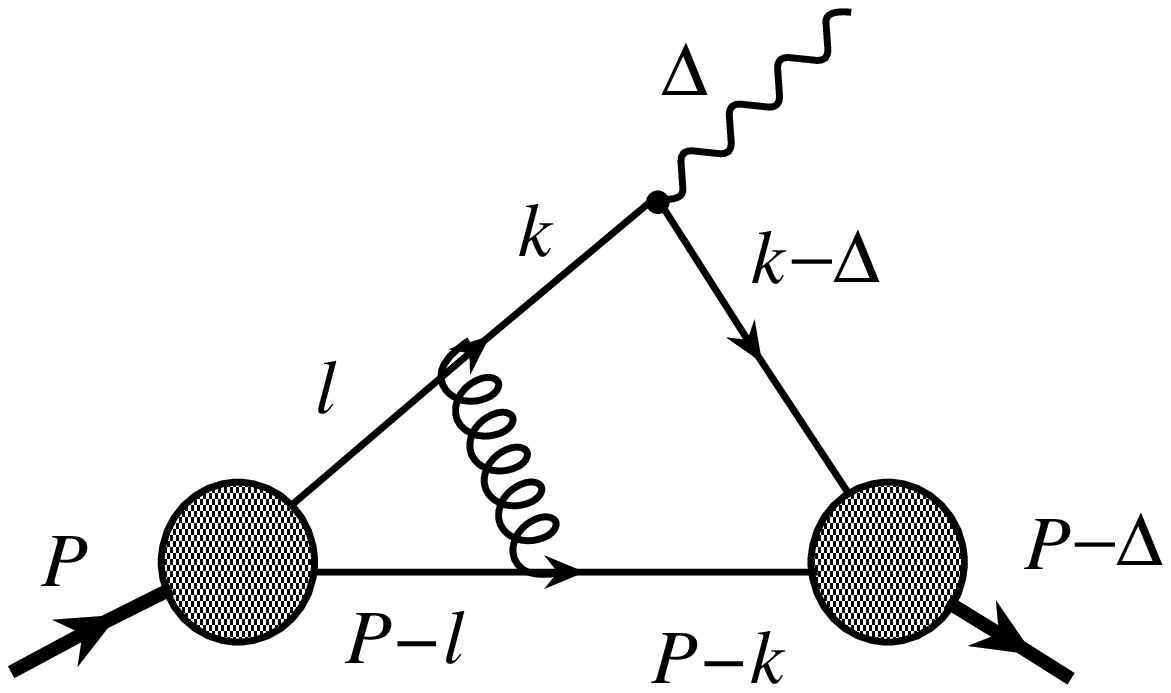}
\end{minipage}
\hspace*{0.25cm}
\begin{minipage}[c]{0.3\hsize}
\epsfig{width=\hsize,file=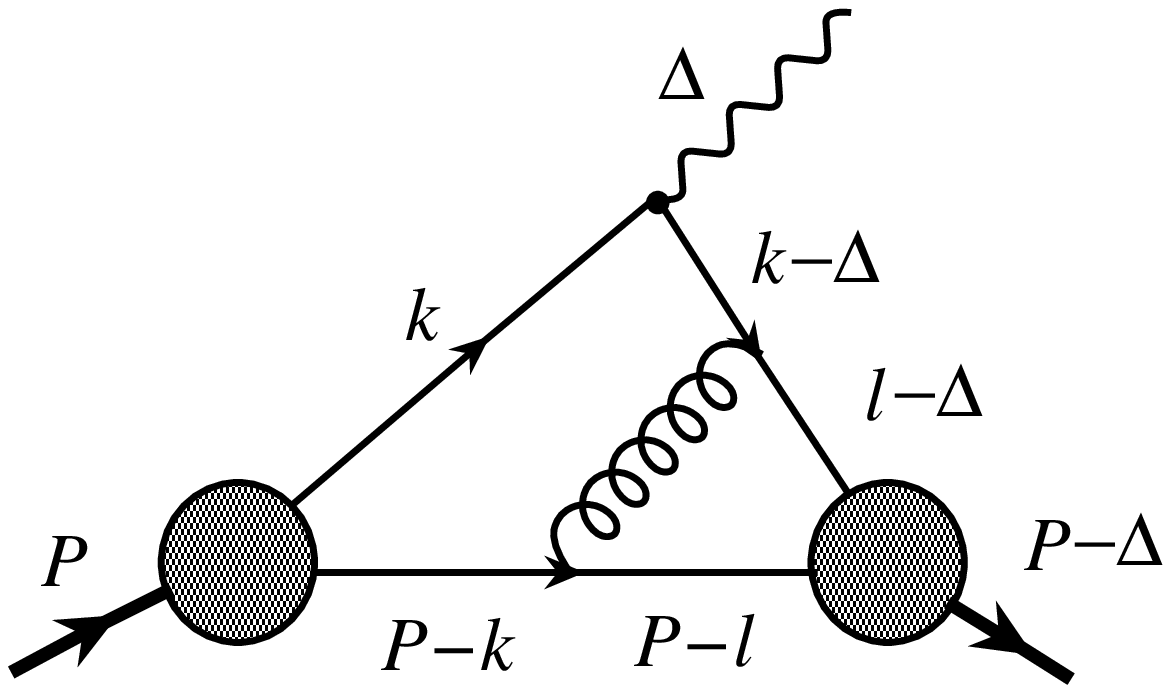}
\end{minipage}
\hspace*{0.25cm}
\begin{minipage}[c]{0.3\hsize}
\epsfig{width=\hsize,file=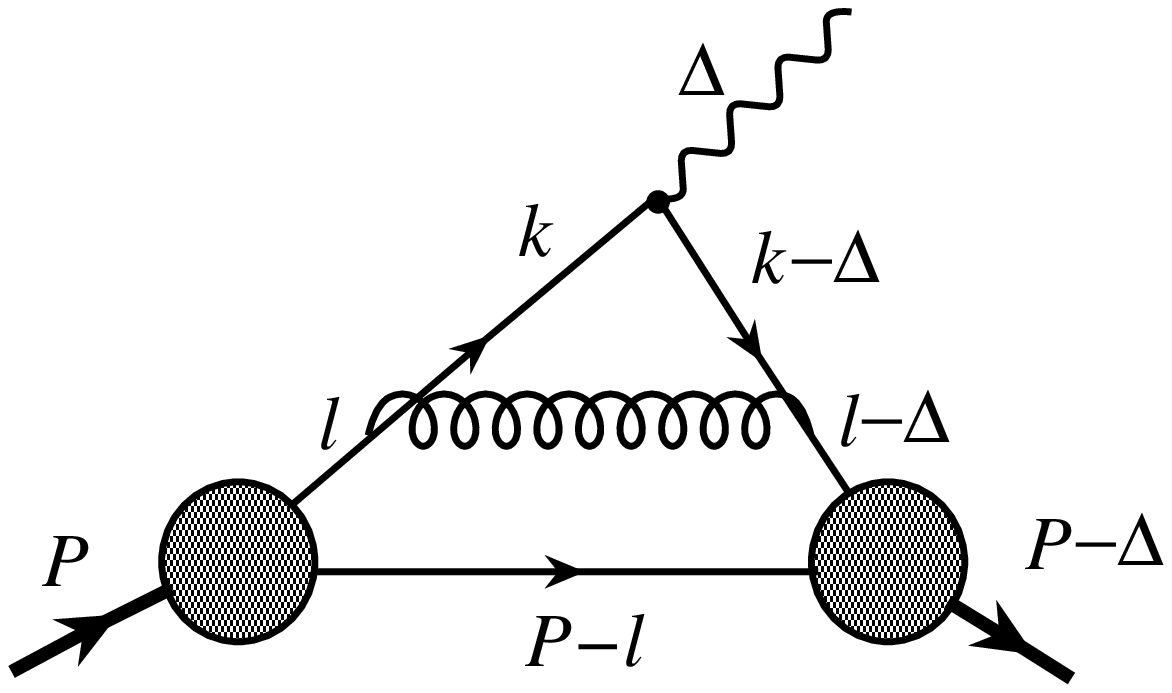}
\end{minipage}\vspace{0.3cm}
\caption{Next-Fock-State corrections to DVCS which we denote S1, S$1'$ 
and S2 (left to right).}
\label{iniDia}
\end{figure}

Given that there had been relatively little progress in finding 
exact, or even model, form of the 3-body LF wavefunction
in QCD from first principles, we choose to model it by relating 3-body
Fock states to
2-body Fock states. As shown in Fig.\ref{23body},  we
link 2-body states and 3-body states via a kernel including,
in principle, all QCD processes which result in the production of one gluon
along with the constituent quarks. Because of  the complexity of
such general kernel,  in this work we limit ourselves to
model it by the lowest-order simplest possible process
shown by  the right diagrams in Fig.\ref{23body}.
Thus, 
when computing the GPD of the pion we will concentrate on the
additional processes described
by the covariant diagrams in Fig.\ref{iniDia}.

\begin{figure}
\centering
\begin{minipage}[c]{0.3\hsize}
\epsfig{width=\hsize,file=jis1_1.ps}
\end{minipage}
\hspace*{0.25cm}
\begin{minipage}[c]{0.3\hsize}
\epsfig{width=\hsize,file=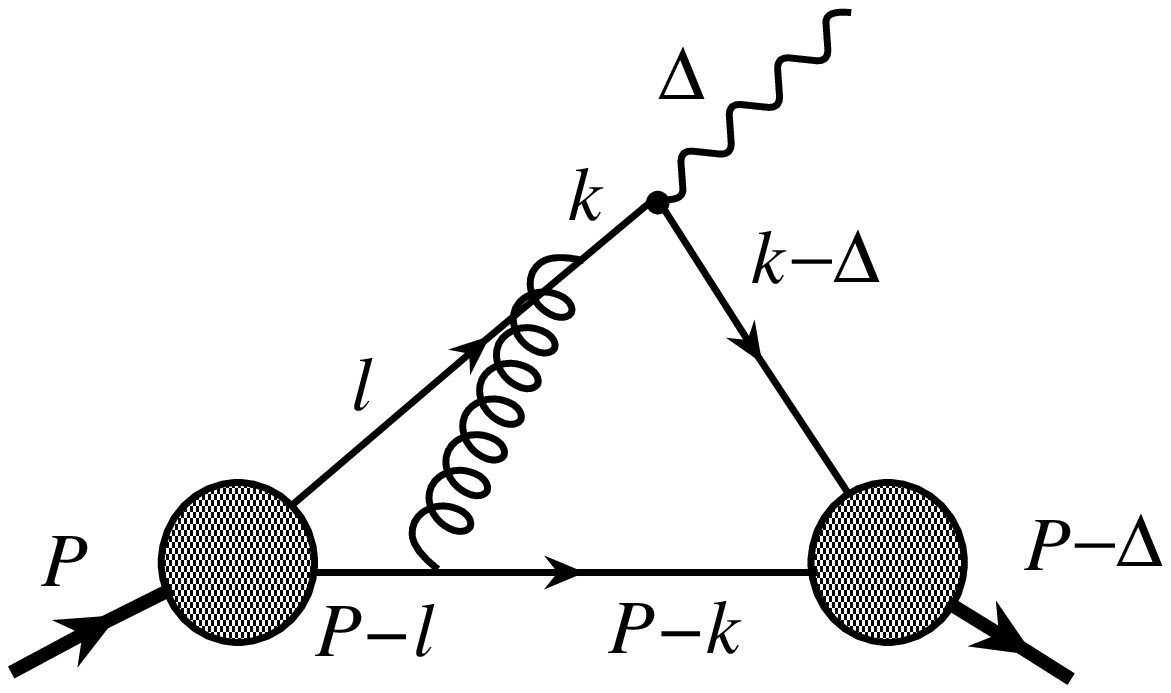}
\end{minipage}
\hspace*{0.25cm}
\begin{minipage}[c]{0.3\hsize}
\epsfig{width=\hsize,file=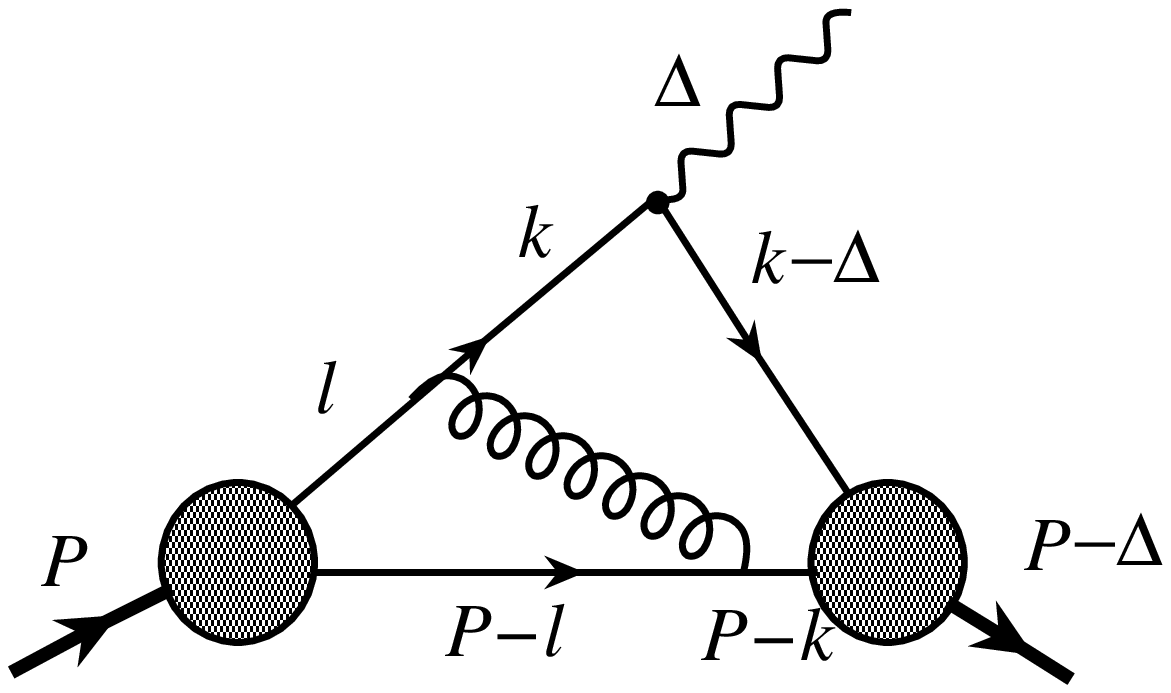}
\end{minipage}\vspace{0.3cm}
\caption{Different time ordered contributions for $x>\zeta$ in S1.}
\label{timeS1}
\end{figure}

In Ref.\cite{Tiburzi:2001je}, it was 
stated
that, when considering processes such as those of 
Fig.\ref{iniDia} in the LF dynamics, the time orderings when the 
additional gluon was exchanged entirely before (after) the photon emission
should be omitted. This is because such
contributions are related  to initial (final) state interactions corresponding
to iteration of the LF wavefunction with BS kernel. Due to BS equation,
these exchanges  can be entirely absorbed back into the LF wavefunction.
For example, the first diagram in Fig.\ref{iniDia}, which we call S1,
yields the  three time-ordered contributions shown in Fig.\ref{timeS1}.
According to Ref.\cite{Tiburzi:2001je}, two left diagrams in Fig.\ref{timeS1}
are iterations of the LF wavefunction with the BS kernel
in the one-gluon exchange approximation
and, thus, should
be absorbed into the LF vertex $h^{\rm val}_{LF}(y,{\bf k}_\perp )$. 
The only remaining contribution is the one where the
gluon is present at the time of the photon emission, as shown by the right 
diagram in Fig.\ref{timeS1}.
Using the LF-perturbation theory, we write this contribution as
\begin{eqnarray}
& &\lefteqn{\int \frac{dk^+d^2{\bf k}_\perp}{16\pi^3}
\frac{dl^+d^2{\bf l}_\perp}{16\pi^3} 
f V h_{\rm in}(l;P) h_{\rm out}(k-\Delta;P-\Delta)}  \nonumber \\ & & \times 
(-\zeta) \times  
{\rm Tr} \,  [\gamma_5 (\dr l -\dr P +m)\gamma^\alpha 
\overline{ (\dr k -\dr P +m)} \gamma_5
(\dr k-\dr \Delta +m)\dr q \overline{ (\dr k+m)}
\gamma_\alpha \overline{(\dr l+m)}]
\nonumber \\ & & \times  
(-1) \times \left [l^+k^+(k-\Delta)^+(P-k)^+(P-l)^+(l-k)^+\right ]^{-1}
 \label{S1LF} \\ 
& & \left \{ 
[P^--l^-_{\rm on}-(P-l)^-_{\rm on}][P^--k^-_{\rm on}-(P-l)^-_{\rm on}-(l-k)^-_{\rm on}]
\nonumber \right. \\ & & \left. 
[P^--\Delta^--(k-\Delta)^-_{\rm on}-(P-l)^-_{\rm on}-(l-k)^-_{\rm on}]
[P^--\Delta^--(k-\Delta)^-_{\rm on}-(P-k)^-_{\rm on}]
\right \}^{-1} \ .\nonumber 
\end{eqnarray}
By the overline we denote the parts which must be taken with 
the instantaneous contribution, {\it i.e.}, with  \mbox{$k^-=P^-_{\rm tot}-\sum k^-_i$.}
This diagram contributes only to the region $\zeta<x<y$,  and is
zero otherwise. See Eq.~(\ref{genform}) below for the definition of $f$ and $V$.

\begin{figure}
\centering
\begin{minipage}[c]{0.3\hsize}
\epsfig{width=\hsize,file=jis1p_1.ps}
\end{minipage}
\hspace*{0.25cm}
\begin{minipage}[c]{0.3\hsize}
\epsfig{width=\hsize,file=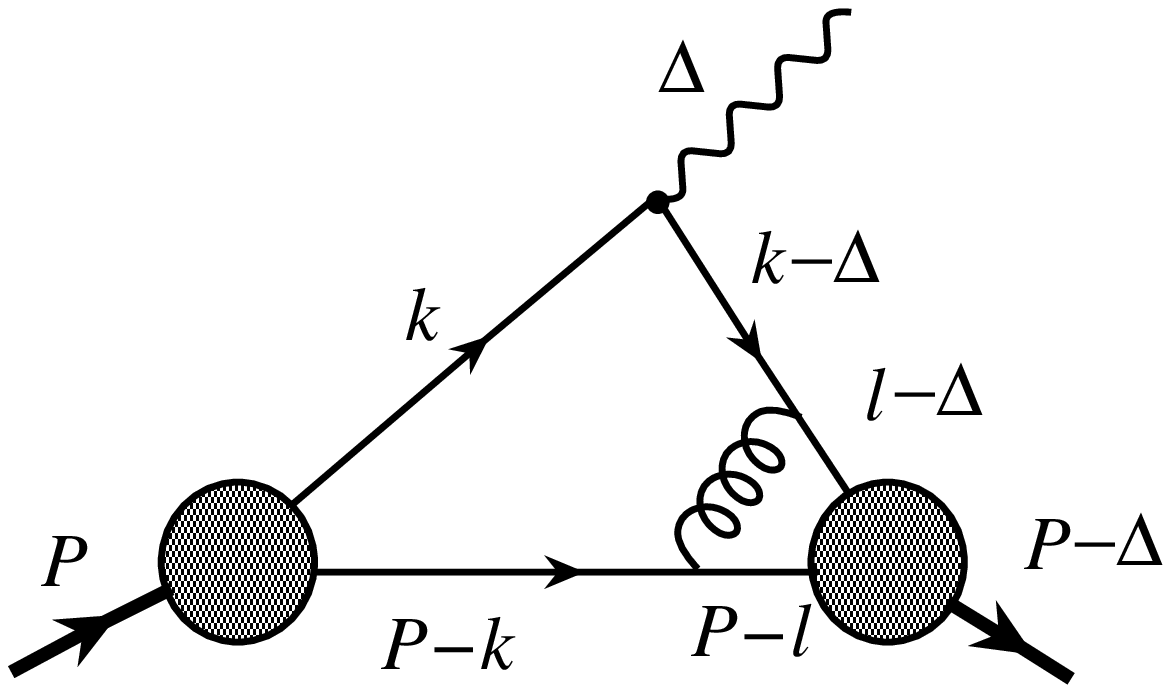}
\end{minipage}
\hspace*{0.25cm}
\begin{minipage}[c]{0.3\hsize}
\epsfig{width=\hsize,file=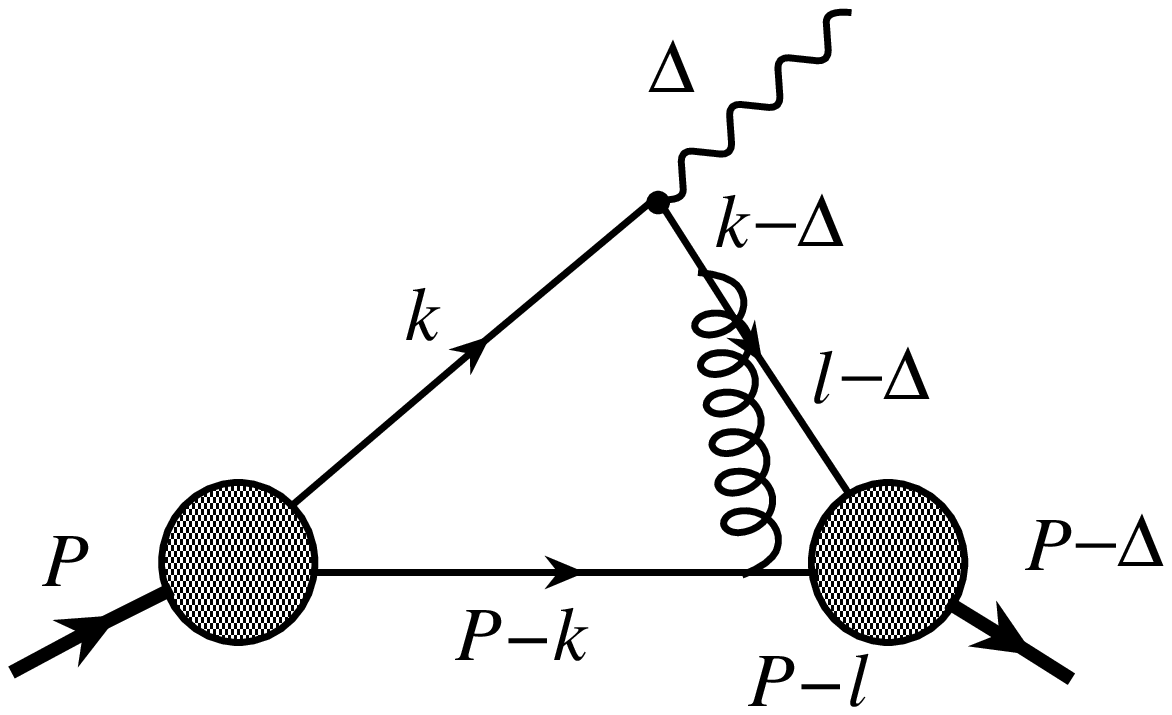}
\end{minipage}\vspace{0.3cm}
\caption{Different time ordered contributions for $x>\zeta$ in S$1'$.}
\label{timeS1p}
\end{figure}

Similarly, for the final state interaction term  S$1'$,  one can get three
time-ordered diagrams in DGLAP region (modulo the instantaneous diagrams) shown 
in Fig.\ref{timeS1p}. According
to Ref.\cite{Tiburzi:2001je},  
out of these three, only the 
left diagram in Fig.\ref{timeS1p} should be kept.
 Using the LF-perturbation theory rules,
we obtain that this contribution is given by 
\begin{eqnarray}
& &\lefteqn{\int \frac{dk^+d^2{\bf k}_\perp}{16\pi^3}
\frac{dl^+d^2{\bf l}_\perp}{16\pi^3} 
f V h_{\rm in}(k;P) h_{\rm out}(l-\Delta;P-\Delta)} \nonumber \\ & &\times
(-\zeta 
Tr[\gamma_5 \overline{(\dr k -\dr P +m)}
\gamma^\alpha (\dr l -\dr P +m) \gamma_5
\overline{(\dr l-\dr \Delta +m)}\gamma_\alpha 
\overline{(\dr k - \dr \Delta +m)}\dr q (\dr k+m)])\nonumber \\ & &\times
(-1)  \times \left [(l-\Delta)^+k^+(k-\Delta)^+(P-k)^+(P-l)^+(l-k)^+ \right ]^{-1} 
\label{S1pLF}\\ & &\left \{ 
[P^--k^-_{\rm on}-(P-k)^-_{\rm on})(P^--k^-_{\rm on}-(P-l)^-_{\rm on}-(l-k)^-_{\rm on}]^{-1} 
 \nonumber  \right. \\ & & \left. 
[P^--\Delta^--(k-\Delta)^-_{\rm on}-(P-l)^-_{\rm on}-(l-k)^-_{\rm on}]
[P^--\Delta^--(l-\Delta)^-_{\rm on}-(P-l)^-_{\rm on}] \right \}^{-1}  \ , \nonumber 
\end{eqnarray}
where the overline has the same meaning as before.
This term only contributes in the DGLAP region $\zeta<x<y$.

\begin{figure}
\centering
\begin{minipage}[c]{0.35\hsize}
\epsfig{width=\hsize,file=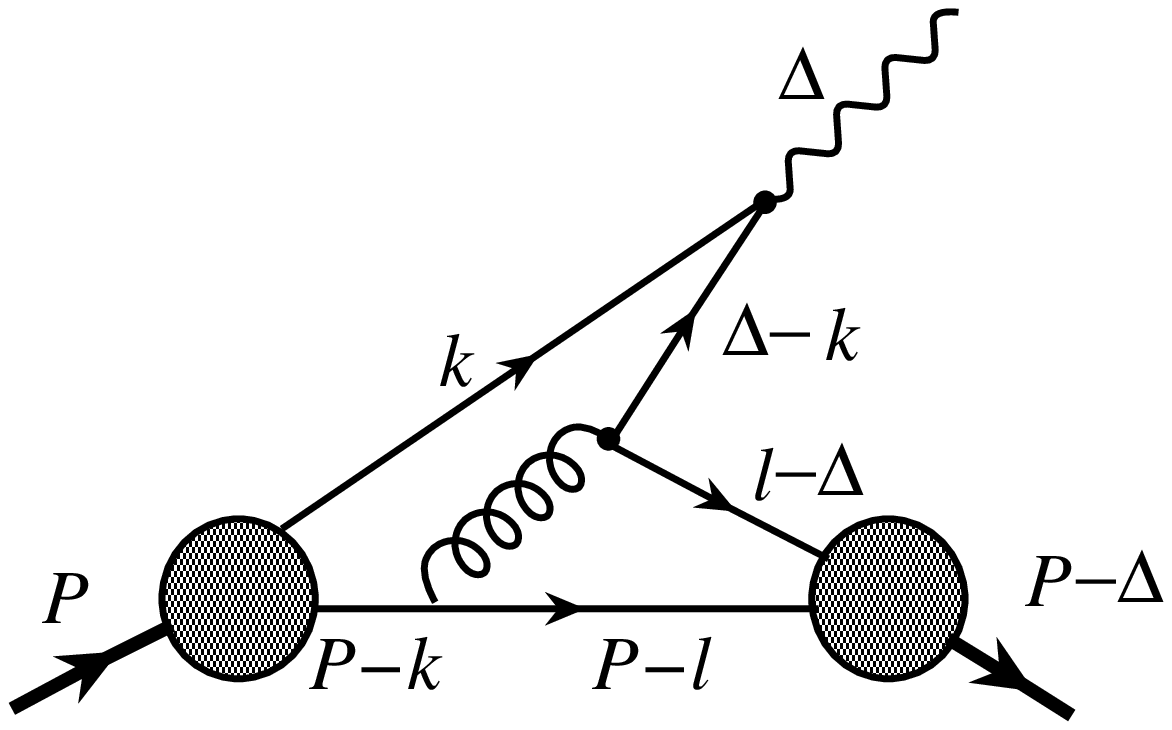}
\end{minipage}
\hspace*{0.35cm}
\begin{minipage}[c]{0.35\hsize}
\epsfig{width=\hsize,file=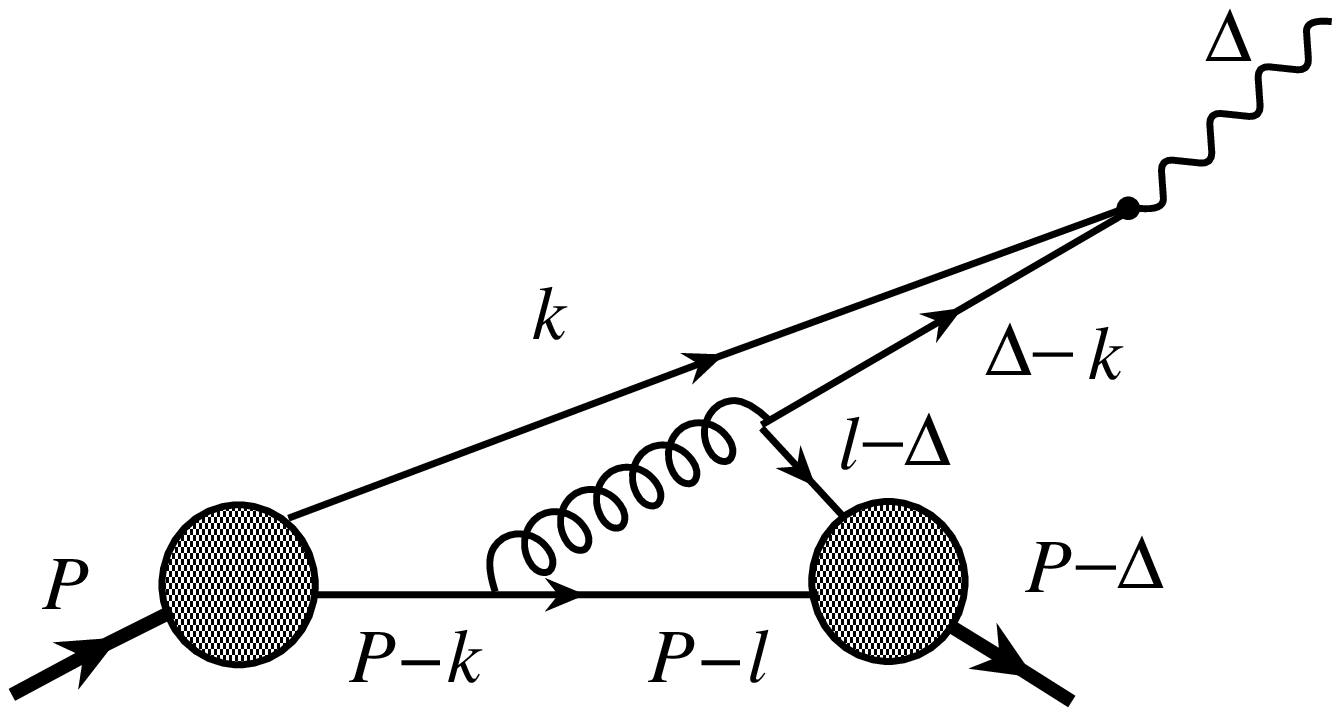}
\end{minipage}\vspace{0.3cm}
\caption{Different time ordered contributions for $x<\zeta$ in S$1'$.}
\label{timeS1pp}
\end{figure}

In the ERBL region,  there are two possible time-ordered 
processes (Fig.\ref{timeS1pp}) which all should be retained as neither of them can
be absorbed into the initial or final state vertex. The sum of these
contributions 
yields the full covariant diagram, and we may use integration by poles, as
described in greater details later, to compute
this case for $x<\zeta<y$. Analogously, for the process
described by the  diagram S2 we keep all time-ordered contributions
as, again,  neither of them is absorbed into initial or final state wavefunction
and we use integration by poles in this case as well.

\begin{figure}
\centering
\begin{minipage}[c]{0.35\hsize}
\epsfig{width=\hsize,file=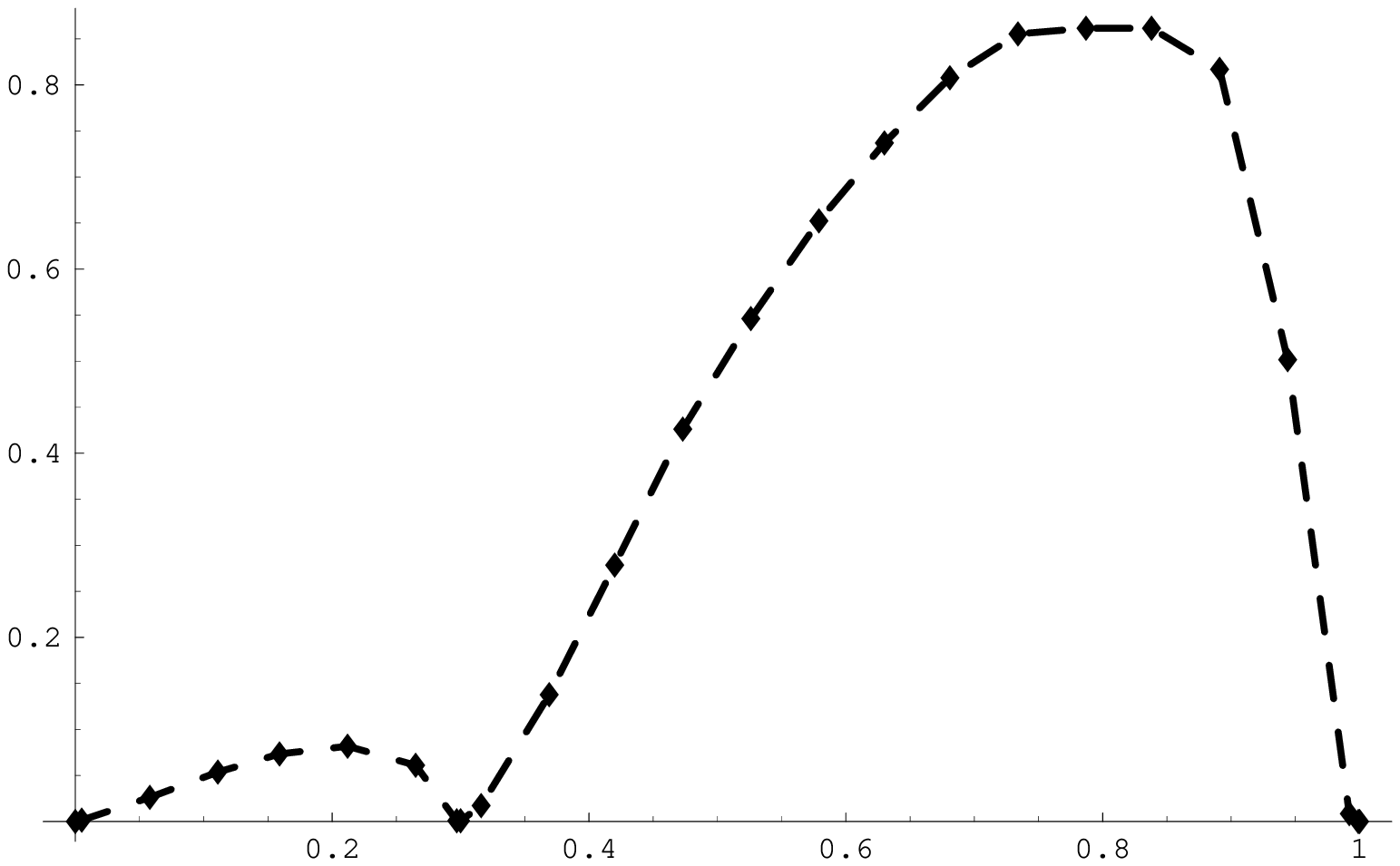}
\end{minipage}
\hspace*{0.25cm}
\begin{minipage}[c]{0.35\hsize}
\epsfig{width=\hsize,file=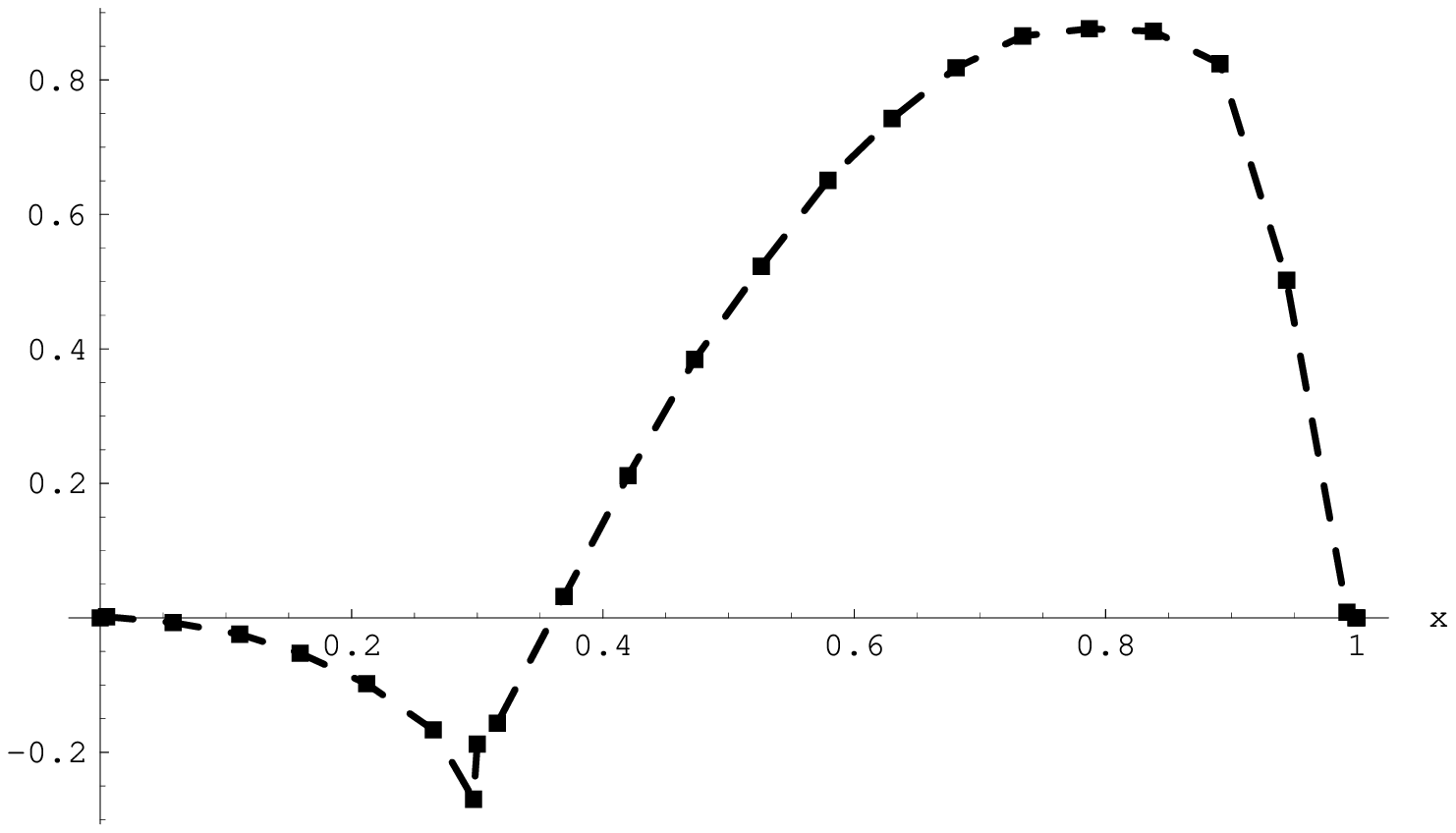}
\end{minipage}\vspace{0.3cm}
\caption{The pion GPD in LFBS approach  with 2-body Fock states only (left),
and with inclusion of  specifically LF-time-ordered $q\bar q g$ contributions (right).
The latter exhibits a discontinuity between DGLAP and ERBL region
near $x=0.3$ ($\zeta=0.3$ and $t=-0.5 $\,GeV$^2$).}
\label{zeta03_1}
\end{figure}

Combining  these contributions,
we calculate the GPD for $\zeta=0.3$ and $t=-0.5 $\,GeV$^2$.
We use $G_\pi=0.32$ and $\alpha_s \approx 0.5$. The results of our
computation are presented in Fig.\ref{zeta03_1}.
As expected, ${\cal F}_\pi(\zeta,\zeta,t)\neq 0$. However, we also notice
a discontinuity in the GPD in Fig.\ref{zeta03_1} (the dashed line) 
near $x=\zeta=0.3$.
The origin of this discontinuity can be attributed to the approximate nature
of the model LF wavefunction that we used in our computation (namely, a Gaussian)
while the 
Statement
of Ref.\cite{Tiburzi:2001je} is only valid for the LF
wavefunctions obtained from the exact solution of the BS equation. This shows that
we cannot forget the time-ordered contributions omitted according to
the prescription of Ref.\cite{Tiburzi:2001je} ({\it i.e.} the first two time-ordered
diagrams in Fig.\ref{timeS1} and the last two diagrams in Fig.\ref{timeS1p})
but should include them all in our calculations.
Taking into account  these contributions 
can be seen as iterating our model LF wavefunction
with the BS kernel once,  to improve the quality of approximate wavefunctions. Of course,
that would have no effect if we would 
  already have the LF wavefunction
as an exact solution of the BS equation in agreement with Ref.\cite{Tiburzi:2001je}.
Since the covariant amplitude is expected to be continuous near $x=\zeta$, 
and because retaining all time-ordered diagrams is equivalent to working
with the covariant amplitudes, including these contributions is crucial
to maintain the continuity of the GPD in the BSLF analysis with the
model LF wavefunctions.

In order to consistently take into account the  higher Fock states in our approach, we thus
consider contributions from all processes
in all time-ordered regions. 
We treat our calculations
as corrections due to $q\bar q g$ Fock states in addition to the ``effective"
expression obtained in the LFBS approach of Ref.\cite{Choi:2001fc,Choi:2002ic}.
Then, each contribution in Fig.\ref{iniDia} 
corresponds to an expression in the covariant form
\begin{equation}\label{genform} \zeta \, 
f \int \frac{dk}{(2\pi)^4}\frac{dl}{(2\pi)^4} V H_{\rm in}
 \, H_{\rm out}\frac{-{\rm Tr} \, \, [\dots]}
{{\rm Denominator}\, [\dots]}
\,  , 
\end{equation}
where the overall coefficient is
$$
f= C_F (-i g_s)^2 i^5(-i)=- C_F g_s^2 \ , 
$$
 the effective
vertex function is 
$$
V=\frac1{\zeta} \, {(-i e_q^2 N_c)(-g^{\mu \nu})} \left (\frac{1}{(q+k)^2-m^2+i \varepsilon}-
\frac{1}{(k-q')^2-m^2+i \varepsilon} \right ) 
$$ 
and the trace is computed for
the diagram with the effective-vertex contributing the $\dr\! q$ factor.
We calculated the contribution from these processes 
using the  LF momentum variables.   First, we  
carried out integration by poles in Eq.~(\ref{genform}). 
In the Cauchy integration over $k^-$ and $l^-$,  we used the poles located
in the opposite halves of the complex $k^-$ and $l^-$ planes so that
always a factor $(2\pi i)$ or $(-2\pi i)$ was introduced. After integration by poles, 
we are left with the expression
\begin{eqnarray}
& &\lefteqn{ \zeta \, f  \int \frac{dk^+d^2{\bf k}_\perp}{16\pi^4}
\frac{dl^+d^2{\bf l}_\perp}{16\pi^4} 
\frac{2\pi i}{2} \frac{-2\pi i}{2} 
V \, h_{LF} h'_{LF}
{\left (-\overline{{\rm Tr} \, \,[\dots]} \right ) }/ 
\left ({\overline{{\rm Denominator}\, [\dots]}} \right )
 } \\ & = &
N_c(-i e_q^2)(-g^{\mu\nu})f\int\limits^1_0 dx \frac{P^+}{\zeta q^-}
\left ( \frac 1{x-\zeta+i\varepsilon} + \frac{1}{x-i\varepsilon} \right )
\int \frac{dyd^2{\bf l}_\perp}{16\pi^3}\frac{d^2{\bf k}_\perp}
{16\pi^3}\, h_{LF} h'_{LF} \, 
 \left ({-\zeta \overline {{\rm Tr} \, \,[\dots]}} \right ) / 
 \left ({\overline{{\rm Denominator}\, [\dots]}} \right ) 
\  ,\nonumber
\end{eqnarray}
where the overline means that the expressions are taken with $k^-$ and $l^-$ 
values corresponding
to  specified  poles for $k^-$ and $l^-$ in the complex plane.
Comparing with $i M^{\uparrow\uparrow}$, $i M^{\downarrow\downarrow}$ in
Eq.~(\ref{CDVCS2}) and noting that
${P^+ Q^2}/{\zeta q^-}\approx 1$ for  DVCS, we identify the corresponding
contribution to ${\cal F}_\pi$ from each process as
\begin{equation}\label{LFGPDx}
f N_c
\int \frac{dyd^2{\bf l}_\perp}{16\pi^3}
\frac{d^2{\bf k}_\perp}{16\pi^3} \,  h_{LF} h'_{LF} \, 
\left ( {-\zeta \overline{{\rm Tr} \, \,[\dots]}/Q^2}\right ) / 
\left ( {\overline{{\rm Denominator}\,[\dots]}} \right )
 \ .
\end{equation}
In this form, the tree-level contribution is simply
\begin{equation}\label{treelevel}
\pm N_c
\int \frac{d^2{\bf k}_\perp}{16\pi^3} \, h_{LF} h'_{LF}  \, 
\left ( {-\zeta \overline{{\rm Tr} \,  \, [\dots]}/Q^2} \right )/
 \left ({\overline{{\rm Denominator}\, [\dots]}} \right ) \ 
,
\end{equation}
and the higher Fock states 
corrections carry additional factor of ${f}/{16\pi^3}$ relative to
the tree-level. The  $\pm$ sign 
in Eq.~(\ref{treelevel}) refers to two possibilities for the pole
selection in the upper or  lower halves of the complex plane, 
as we discussed in Section II.
Each LF contribution can be constructed using Eq.~(\ref{LFGPDx}) and the
pole assignment for $k^-$ and $l^-$ presented in Appendix \ref{app1}.

For each particular 
diagram in Fig.\ref{iniDia},
we  obtain the following expressions. \\
{\bf (S0)} The diagram S0 in Fig.\ref{S0} corresponds to the
covariant expression
\begin{equation}
\int \frac{dk}{(2\pi)^4} N_c V 
\frac{(-\zeta){\rm Tr} \,  \, [\gamma_5 (\dr k -\dr P +m)\gamma_5 (\dr k - \dr \Delta +m) \dr q (\dr k + m)]}
{(k^2-m^2+i\varepsilon)((P-k)^2-m^2+i \varepsilon)((k-\Delta)^2-m^2+i\varepsilon)} \ 
\times H_{\rm in}(k;P) H_{\rm out}(k-\Delta, P-\Delta) \ ,
\label{S0cov} 
\end{equation}
so that the LF-expression is given by
\begin{eqnarray}
\pm\int \frac{dk}{16\pi^3} N_c 
\frac{(-\zeta)\, {\rm  Tr} \, [\gamma_5 (\dr k -\dr P +m)\gamma_5 (\dr k - \dr \Delta +m) \dr q (\dr k + m)]}
{(k^2-m^2+i\varepsilon)((P-k)^2-m^2+i \varepsilon)((k-\Delta)^2-m^2+i\varepsilon)} h_{LF} h'_{LF}
\label{eqn001} \ , 
\end{eqnarray}
where $k^-$ is set to its value at the corresponding pole. 
This case was in detail considered in the previous section.
The $\pm$ sign in Eq.~(\ref{eqn001}) shall be read as plus for $x>\zeta$ and
minus for $x<\zeta$. \newline
{\bf (S1)} The initial-state-interaction contribution 
S1 in the left diagram of  Fig.\ref{iniDia} corresponds to the following
expression
\begin{eqnarray}
& &\lefteqn{\int \frac{dk}{(2\pi)^4}\frac{dl}{(2\pi)^4} f V H_{\rm in}(l;P) 
H_{\rm out}(k-\Delta;P-\Delta)}  \\ & \times &
\frac{(-\zeta)  \, 
{\rm Tr} \,  [\gamma_5 (\dr l -\dr P +m)\gamma^\alpha (\dr k -\dr P +m) \gamma_5
(\dr k-\dr\! \Delta +m)\dr q (\dr k+m)\gamma_\alpha (\dr l+m)]}
{((P-l)^2-m^2+i\varepsilon)((P-k)^2-m^2+i\varepsilon)
((k-\Delta)^2-m^2+i\varepsilon)(k^2-m^2+i\varepsilon)(l^2-m^2+i\varepsilon)
((l-k)^2+i\varepsilon)} \ . \nonumber
\label{S1cov}
\end{eqnarray}
Here, as well, $k^-$ and $l^-$ should be taken at the corresponding pole as
presented in Appendix \ref{app1}.\newline
{\bf (S1$^\prime$)} The final-state-interaction diagram S$1'$ in the center
diagram of Fig.\ref{iniDia} corresponds to 
\begin{eqnarray}
& &\lefteqn{\int \frac{dk}{(2\pi)^4}\frac{dl}{(2\pi)^4} f V H_{\rm in}(k;P) 
H_{\rm out}(l-\Delta;P-\Delta) \label{S1pcov} }\\ & \times &
\frac{ (-\zeta)\,  
{\rm Tr} \,  \, [\gamma_5 (\dr k -\dr\! P +m)\gamma^\alpha (\not l -\dr P +m) \gamma_5
(\dr l-\dr\! \Delta +m)\gamma_\alpha (\dr k - \dr\! \Delta +m)\dr q (\dr k+m)]}
{((P-k)^2-m^2+i\varepsilon)((P-l)^2-m^2+i\varepsilon)
((l-\Delta)^2-m^2+i\varepsilon)((k-\Delta)^2-m^2+i\varepsilon)(k^2-m^2+i\varepsilon)
((l-k)^2+i\varepsilon)}  \ ,
\nonumber
\end{eqnarray}
with the specific values for $k^-$ and $l^-$ obtained from the poles presented in 
Appendix \ref{app1}. \newline
{\bf(S2)} The box-diagram S2 in the right diagram of  Fig.\ref{iniDia} 
is given by 
\begin{eqnarray}
& &\lefteqn{\int \frac{dk}{(2\pi)^4}\frac{dl}{(2\pi)^4} f V H_{\rm in}(l;P)
 H_{\rm out}(l-\Delta;P-\Delta)}\label{S2cov}
 \\ & \times &
\frac{(-\zeta )\, 
{\rm Tr} \,  \, [\gamma_5 (\dr l -\dr P +m)\gamma_5(\dr l -\dr \Delta +m)\gamma^\alpha
(\dr k -\dr \Delta +m)\dr q (\dr k +m)\gamma_\alpha (\dr l+m)]}
{((P-l)^2-m^2+i\varepsilon)(l^2-m^2+i\varepsilon)
((l-\Delta)^2-m^2+i\varepsilon)((k-\Delta)^2-m^2+i\varepsilon)(k^2-m^2+i\varepsilon)
((l-k)^2+i\varepsilon)}\ ,\nonumber 
\end{eqnarray}
with the specific values for $k^-$ and $l^-$ obtained from the poles
listed in Appendix \ref{app1}.
Given the   complexity of the LF expressions obtained after relevant poles substitution, 
we do not 
find it possible to present them more explicitly. The calculations for the pion GPD,
including large portion of symbolic math, were further carried out numerically.

Now, let us discuss our numerical results for the pion GPD with the 
$q\bar q g$ contributions included.  
We performed our calculations for $\zeta=0.1,0.3,0.5$ and $\zeta=0.7$ 
and the value of $-t=0.5$\,GeV$^2$. 
We used $\alpha_s\approx 0.5$ and $G_\pi=0.32$ \cite{Choi:2002ic}. 
While $G_\pi$ may be thought to depend on the values of
$\zeta$ in order to accommodate the
sum rule for the GPD
\begin{equation}
\int {dx} \, {\cal F}(\zeta,x,t) = (1-\zeta / 2) \, F_\pi(t),
\end{equation}
we find, similar to Ref.\cite{Choi:2002ic}, that this condition is satisfied
within $10\%$ for all considered values of $\zeta$ even with $G_\pi$
kept as a constant $G_\pi = 0.32$.
Also, due to the additional
contributions from 2-body and 3-body Fock states and the necessity
to satisfy 
the form-factor normalization condition
\begin{equation}\label{normX}
\int dx \, {\cal F}(\zeta=0,x,t=0) = 1,
\end{equation}
we need to adjust the normalization factor for our model 2-body wavefunctions
introduced in Eq.~(\ref{radial}). We find that a factor of $\chi \approx 1.5$
is needed for the 2-body normalization
to comply with  Eq.~(\ref{normX}) after $q\bar q g$ corrections
are taken into account.

 
\begin{figure}
\centering
\begin{minipage}[c]{0.35\hsize}
\epsfig{width=\hsize,file=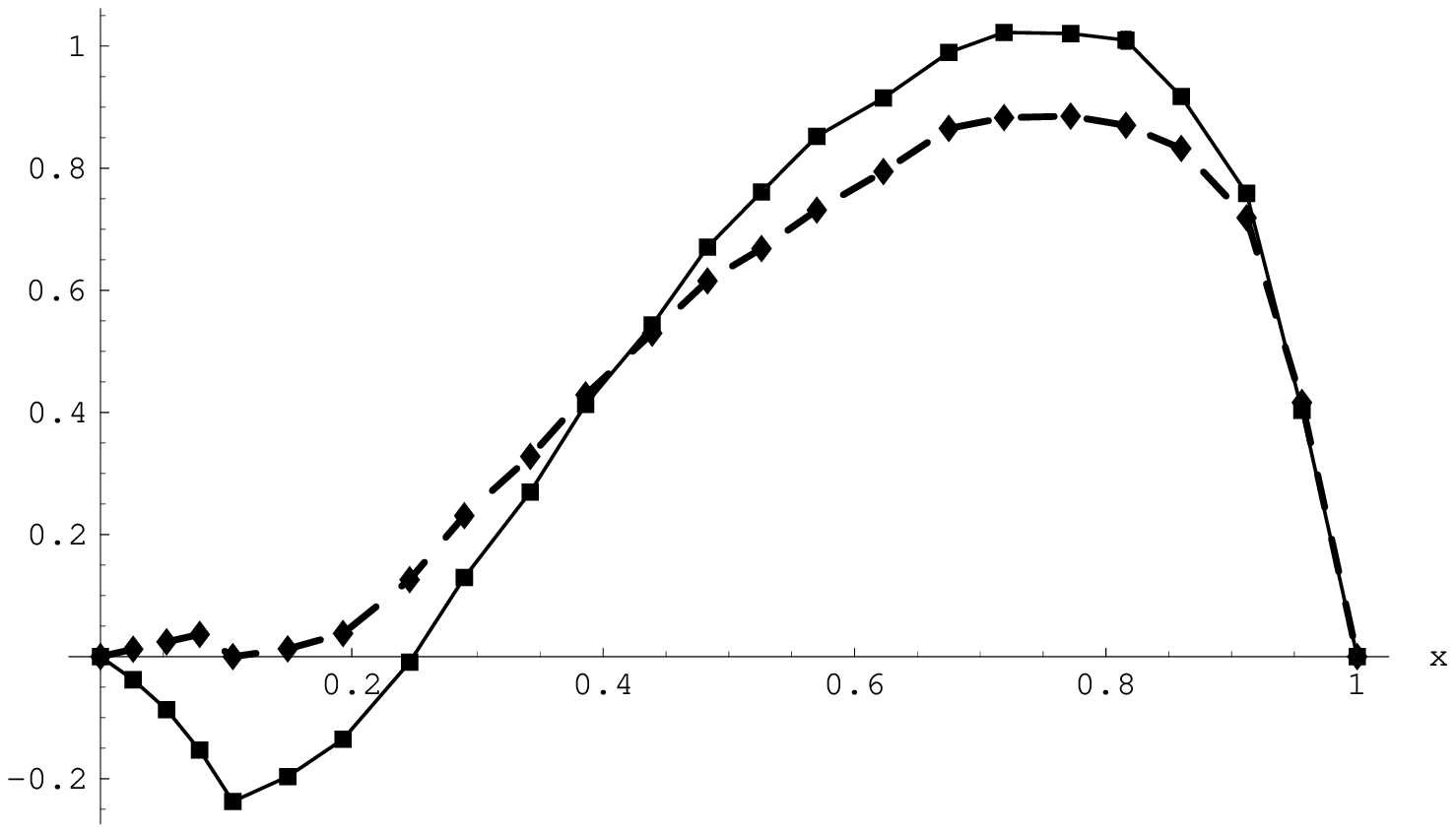}
\end{minipage}
\hspace*{0.25cm}
\begin{minipage}[c]{0.35\hsize}
\epsfig{width=\hsize,file=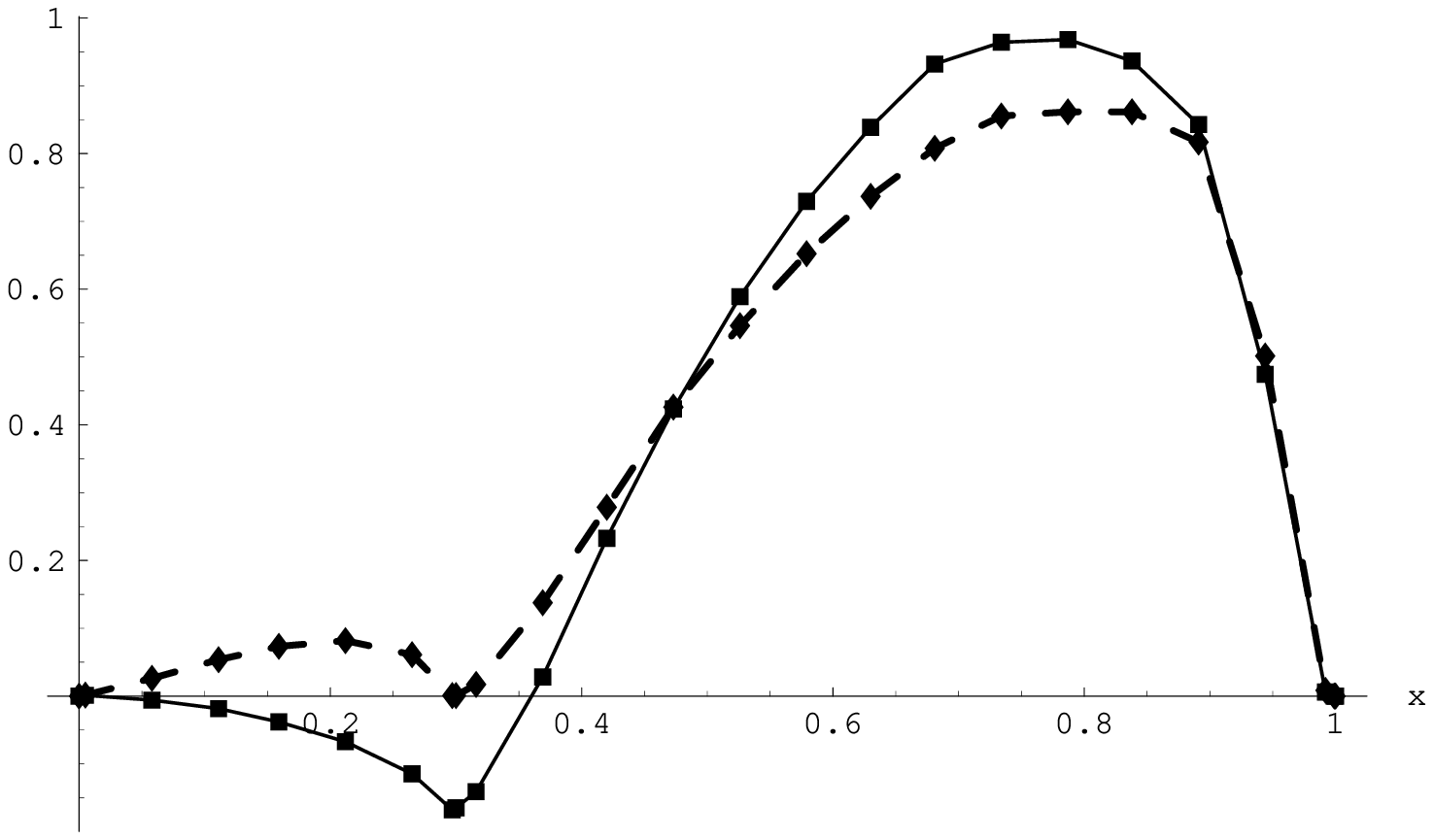}
\end{minipage}
\begin{minipage}[c]{0.35\hsize}
\epsfig{width=\hsize,file=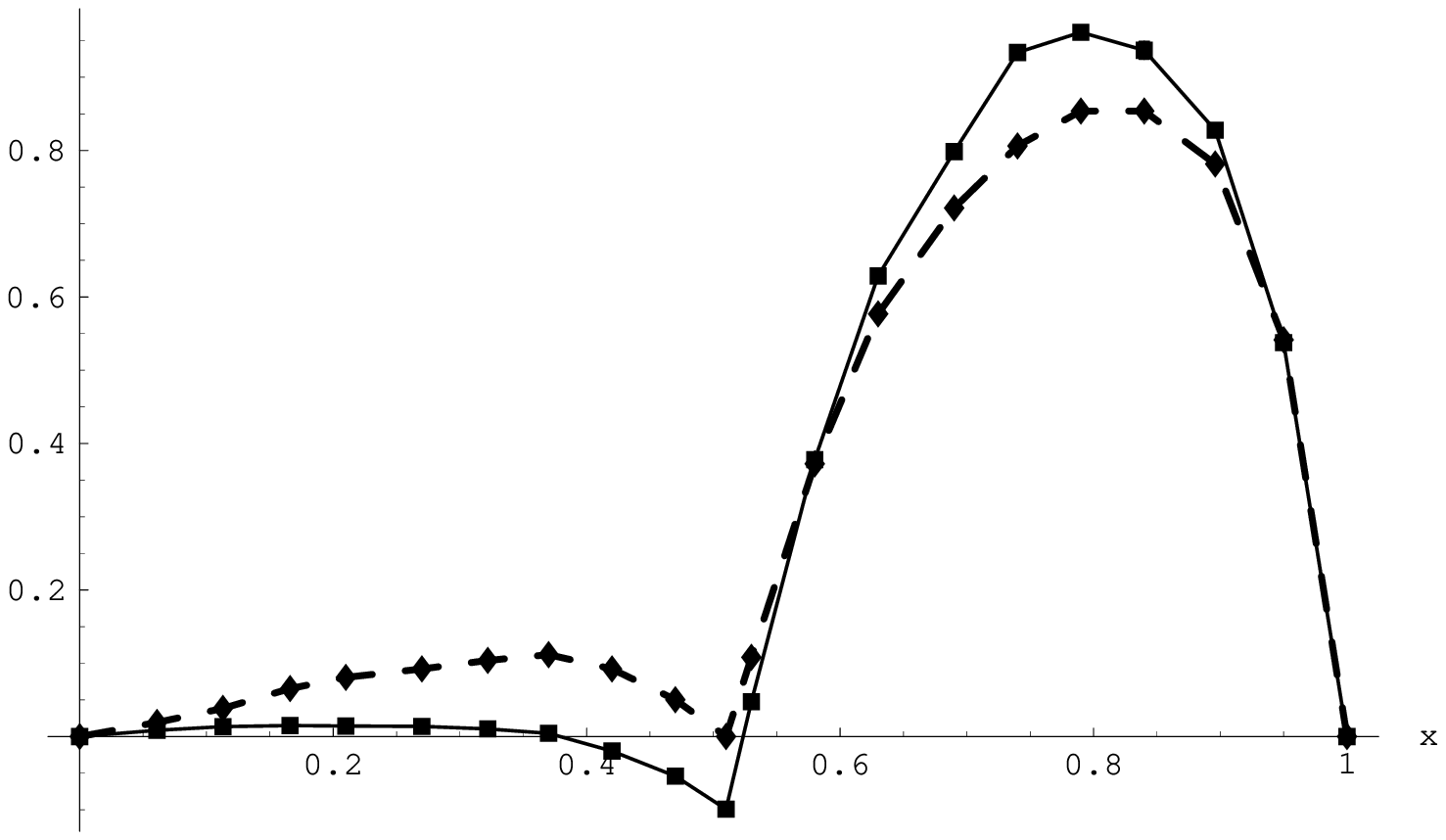}
\end{minipage}
\hspace*{0.25cm}
\begin{minipage}[c]{0.35\hsize}
\epsfig{width=\hsize,file=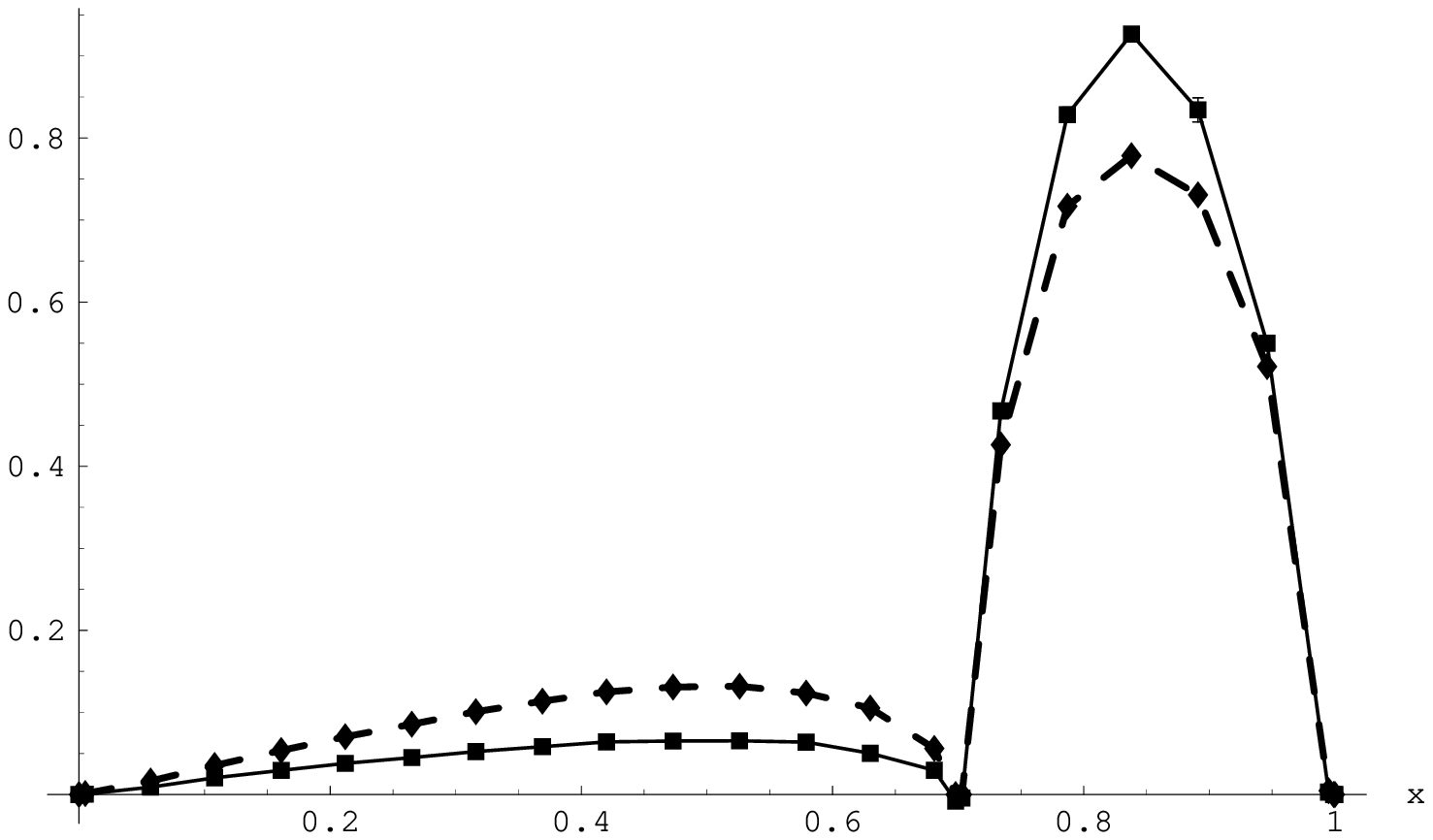}
\end{minipage}\vspace{0.3cm}
\caption{The pion GPD in LFBS approach including
$q\bar q g$ corrections for $-t=0.5$\,GeV$^2$
and different values of $\zeta$. In the upper left, $\zeta=0.1$, while 
$\zeta=0.3$ in the upper right,  $\zeta=0.5$ in the lower left, and
$\zeta=0.7$ in the lower right panel, respectively.
The thick dashed line represents  the GPD calculation in the LFBS approach with
only 2-body Fock states contribution,  and the 
continuous line is the result  for  GPD after  the 
$q\bar q g$ corrections were taken  into account.}
\label{final-gpd}
\end{figure}

As expected, due to the higher Fock state contributions the value of the GPD
at the crossover $x=\zeta$ is nonzero (see Fig.\ref{final-gpd}). Also, we now find that
the GPD is continuous over the entire range of $x$ and $\zeta$ including
$x\approx \zeta$. 
It is crucial to take into account all possible time-ordered
contributions including those that could be formally absorbed into the BS amplitude of
the initial (final) vertex. After  this was done, the connection
between DGLAP and ERBL region is now continuous.

We observed that for the smaller $\zeta$
the higher Fock states introduce dominantly negative corrections in 
ERBL region so that ${\cal F}(\zeta,\zeta,t)$ may become negative. 
Generally, we found
that the GPD has sign-alternating structure, unlike the
effective LFBS treatment result \cite{Choi:2002ic} and
simple DD-based models \cite{Radyushkin:1998es,Mukherjee:2002gb}, 
with nonvalence contribution $x<\zeta$ being often negative
and valence contribution $x>\zeta$ usually positive,
similarly to the results of the model in Ref.\cite{Tiburzi:2003ja}. 
The point ${\cal F}(x,\zeta,t)=0$ in our calculations always gets
shifted from $x=\zeta$ toward the larger values of $x$.
For the larger $\zeta$, however, the structure of the GPD changes and 
the GPD  becomes 
positive for all $x$
with no point at which ${\cal F}(x,\zeta,t)=0$. We also find that the nonvalence
contribution gets suppressed for the larger values of $\zeta$ relative to the
case of small $\zeta$.
The positive valence part of the GPD becomes the dominant contribution in this case.

\section{Conclusion}\label{sec5}
We have taken into account 
 the higher Fock state ($q\bar q g$) contributions to the pion
GPD,  and verified that the value of the GPD at the crossover point $x=\zeta$ is
indeed nonzero ${\cal F}_\pi (\zeta,\zeta,t)\neq 0$.
First, we  followed the 
statement
of Ref.\cite{Tiburzi:2001je} 
and found that, although
${\cal F}_\pi (\zeta,\zeta,t)\neq 0$, a discontinuity near $x=\zeta$ occurs
due to the approximate nature of our model LF wavefunction. The explanation is that 
the 
statement
of Ref.\cite{Tiburzi:2001je} is valid only for the LF wavefunctions obtained from
the exact solution of the BS equation. 
 Thus, we  had to retain the first two time-ordered 
diagrams in Fig.\ref{timeS1} and the last two diagrams in Fig.\ref{timeS1p}. 
Taking into account  these contributions is equivalent
to iterating our model wavefunction with the BS kernel, 
and it is crucial to  maintain the continuity of the GPD in our LFQM analysis.
We thus included all the possible time-ordered diagrams equivalent to the
$q\bar q g$ Fock state contributions shown in Fig.\ref{iniDia}.
We carried out integration by poles using the pole assignment summarized
in Appendix \ref{app1} and numerically computed the pion GPD including
$q\bar q g$ contributions.
From this calculation, we found that the GPD is continuous and the value at
the crossover point $x=\zeta$ is nonzero as expected. The essential finding of the
paper,  namely, the 
link between the nonzero GPD at the crossover point and the higher Fock-state
contributions is not specific to the pion case but 
applicable also for the proton as well as other bound states.

\begin{acknowledgements} 
This work was supported in part by the SURA/JLAB 
Fellowship for Yuriy Mishchenko
and by the grant DE-FG02-96ER 40947 from the U.S. Department of Energy. 
The work of  A.R. was supported by the US 
Department of Energy  contract
DE-AC05-84ER40150 under which the Southeastern
Universities Research Association (SURA)
operates the Thomas Jefferson Accelerator Facility.  
The National Energy 
Research Scientific Computing Center 
is also acknowledged for the grant of computing 
time.

\end{acknowledgements}

\begin{appendix}

\section{Pole locations for the $q\bar q g$ contributions }\label{app1}

(S1) For the initial-state-interaction diagram S1 (left in Fig.\ref{iniDia}),
depending on relative magnitude of ($x$, $y$, $\zeta$) there are six different
kinematic domains. In each of them poles for 
Cauchy integration are chosen as follows.

\vspace{0.5cm}

\begin{tabular}{|r|l|}
\hline
Region & Pole locations \\
\hline
$y<x<\zeta$ & ($l^2-m^2=0$, $(P-k)^2-m^2=0$) \& ($l^2-m^2=0$, $(k-\Delta)^2-m^2=0$) \\
$y<\zeta<x$ & ($l^2-m^2=0$, $(P-k)^2-m^2=0$)  \\
$\zeta<y<x$ & ($l^2-m^2=0$, $(P-k)^2-m^2=0$)  \\
$x<y<\zeta$ & ($(P-l)^2-m^2=0$, $k^2-m^2=0$)  \\
$x<\zeta<y$ & ($(P-l)^2-m^2=0$, $k^2-m^2=0$)  \\
$\zeta<x<y$ & ($(P-l)^2-m^2=0$, $k^2-m^2=0$) \& ($(P-l)^2-m^2=0$, $(k-\Delta)^2-m^2=0$) \\
\hline
\end{tabular}

\vspace{0.5cm}

(S$1'$) The final-state-interaction diagram S$1'$ (center in Fig.\ref{iniDia}).
In each kinematic domain the poles are chosen as follows.

\vspace{0.5cm}

\begin{tabular}{|r|l|}
\hline
Region & Pole locations \\
\hline
$y<x<\zeta$ & all poles are in one half-plane, integral is zero \\
$y<\zeta<x$ & all poles are in one half-plane, integral is zero  \\
$\zeta<y<x$ & ($(l-\Delta)^2-m^2=0$, $(P-k)^2-m^2=0$)  \\
$x<y<\zeta$ & ($(P-l)^2-m^2=0$, $k^2-m^2=0$) \& ($(l-\Delta)^2-m^2=0$, $k^2-m^2=0$)  \\
$x<\zeta<y$ & ($(P-l)^2-m^2=0$, $k^2-m^2=0$)  \\
$\zeta<x<y$ & ($(P-l)^2-m^2=0$, $k^2-m^2=0$) \& ($(P-l)^2-m^2=0$, $(k-\Delta)^2-m^2=0$) \\
\hline
\end{tabular}

\vspace{0.5cm}

(S2) The box-diagram S2 (right in Fig.\ref{iniDia}).
Poles are chosen as follows.

\vspace{0.5cm}

\begin{tabular}{|r|l|}
\hline
Region & Pole locations \\
\hline
$y<x<\zeta$ & ($l^2-m^2=0$, $(k-\Delta)^2-m^2=0$) \\
$y<\zeta<x$ & all poles are in one half-plane, integral is zero  \\
$\zeta<y<x$ & all poles are in one half-plane, integral is zero  \\
$x<y<\zeta$ & ($(P-l)^2-m^2=0$, $k^2-m^2=0$) \& ($(l-\Delta)^2-m^2=0$, $k^2-m^2=0$)  \\
$x<\zeta<y$ & ($(P-l)^2-m^2=0$, $k^2-m^2=0$)  \\
$\zeta<x<y$ & ($(P-l)^2-m^2=0$, $k^2-m^2=0$) \& ($(P-l)^2-m^2=0$, $(k-\Delta)^2-m^2=0$) \\
\hline
\end{tabular}

\section{Organization of numerical calculations}\label{app2}

The calculation of  corrections to DVCS originating 
from addition of the  $q\bar q g$ Fock states 
has been performed with the help of  Mathematica program. 
Each contribution was specified by its covariant expression for the trace
obtained from the diagrams in Fig.\ref{iniDia}, the list of 
the denominator factors entering into   Eq.~(\ref{genform}), 
and the assignment of poles for each of 6 possible kinematic domains. 
As a  result,  the  expressions for full amplitudes were constructed.  
Essentially, for each of the diagrams the expression like
the following was generated:
 
\begin{equation}\label{LFGPD0}
{\cal M}=  \frac{(-\zeta)\, \overline{{\rm Tr} \,  \, [\dots]}/Q^2}{\overline{{\rm 
Denominator}[\dots]}} \, 
h_{LF} h'_{LF} \ ,
\end{equation}
where we used for $h_{LF}$
\be\label{lab1}
h_{LF}=\frac{M^2-M_0^2}{\sqrt{M_0}} \, \phi (x, {\bf k}_\perp).
\ee
The wavefunction $\phi(x,{\bf k}_\perp)$ is  given by Eq.~(\ref{radial}).
For  the LF wavefunction of the gauge-boson, 
we used the expression that is 
slightly different from  Eq.~(\ref{Gradial}). Specifically, 
we dropped $m^2$ from the argument of the exponent in ${\bf k''} _\perp ^2+m^2$.
This resulted in a slight shift of the nonvalence contribution in the direction of
the larger $x$ compared to Ref.\cite{Choi:2001fc}. 
However, this effect was rather 
insignificant.

\begin{figure}
\centering
\begin{minipage}[c]{0.3\hsize}
\epsfig{width=\hsize,file=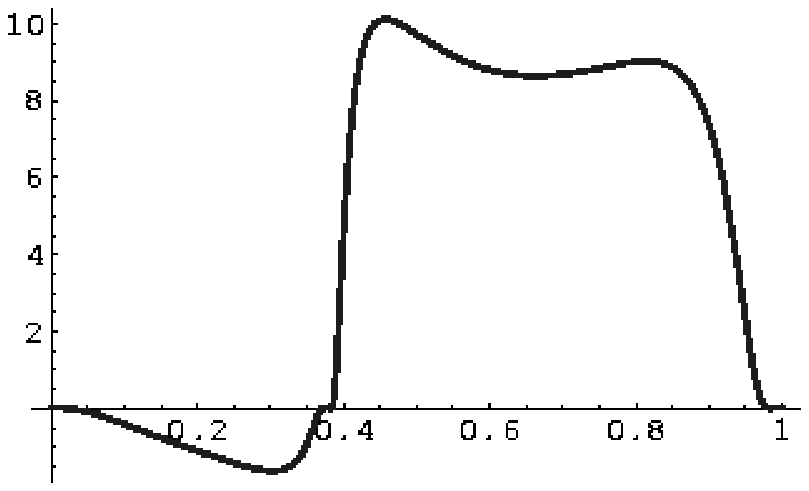}
\end{minipage}
\hspace*{0.25cm}
\begin{minipage}[c]{0.3\hsize}
\epsfig{width=\hsize,file=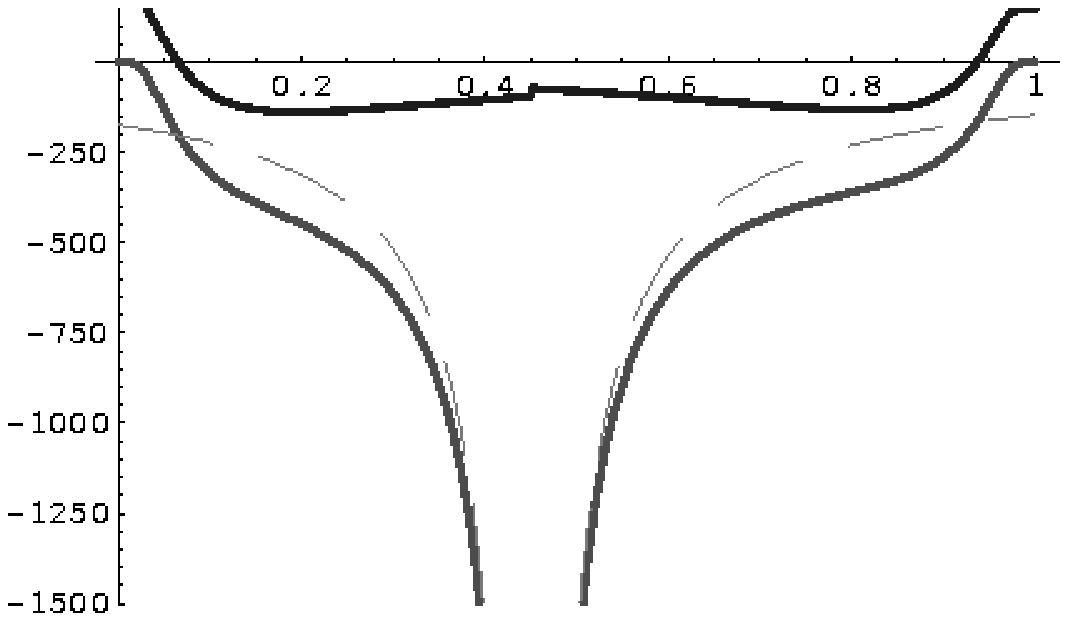}
\end{minipage}
\hspace*{0.25cm}
\begin{minipage}[c]{0.3\hsize}
\epsfig{width=\hsize,file=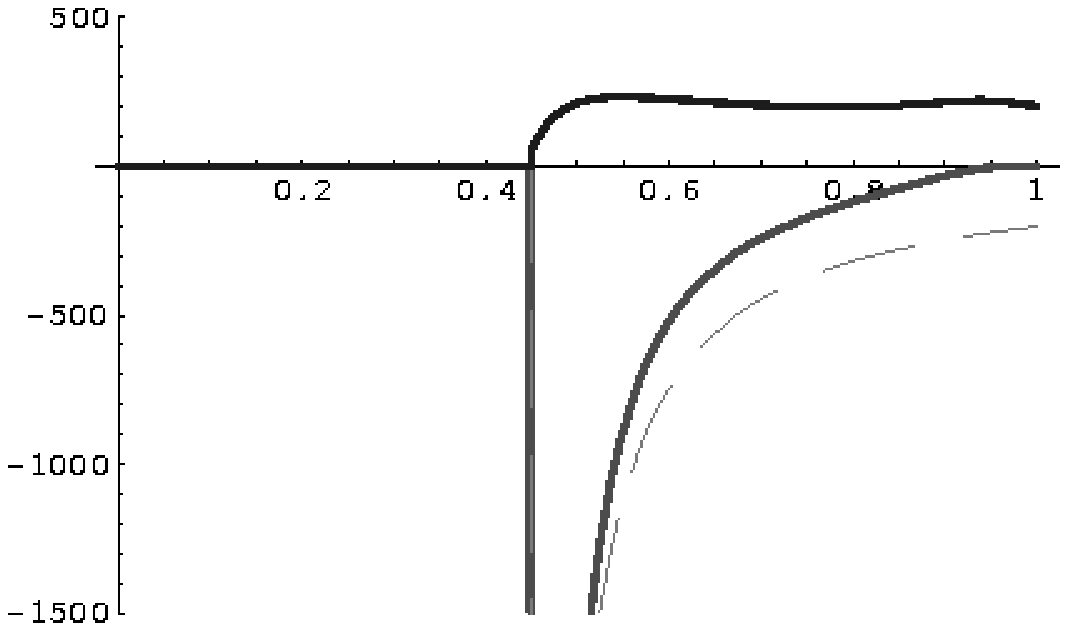}
\end{minipage}\vspace{0.3cm}
\caption{Examples of amplitudes S0, S1, S2 
for some $x$ and $\zeta$ plotted as a function of $y$. Gray line is the
original amplitude and dashed line is the IR-piece subtracted from it.
IR piece is partially integrated analytically. The remaining
amplitude (black) is integrated numerically in 5D using MC method.}
\label{diags01}
\end{figure}

The amplitudes obtained in this way need to be integrated in 5
dimensions, {\it i.e.} $d^2{\bf k}_\perp dy d^2{\bf l}_\perp$. 
This can be  done using 
a Monte-Carlo(MC) algorithm. 
To improve the efficiency of the MC integration, we analyzed 
and
subtracted possible singularities from the amplitudes.
In particular, almost each amplitude carried an IR-singularity  at $y=x$, 
${\bf l}_\perp={\bf k}_\perp$.
While these singularities  are  integrable, 
\be
\int\limits_0 dy \int\limits_{|l_\perp| \leq \Lambda} d^2{\bf l}_\perp \, 
\frac{1}{A (y-x) - ({\bf l}_\perp - {\bf k}_\perp)^2}=\, 
$finite,$
\ee
they  can  seriously degrade the efficiency of the multi-dimensional integration. 
To avoid this, we picked out such
contributions in the form
\begin{equation}\label{LFGPDX}
{\cal M}=\left. \frac{(-\zeta ) \, \overline{{\rm Tr} \, \, [\dots]}/Q^2}{\overline{{\rm Denominator}[\dots]}} \, 
h_{LF} h'_{LF} \right |_{y=x, \,{\bf  l}_\perp={\bf k}_\perp}  \times 
\frac{1}{A (y-x) - ({\bf l}_\perp - {\bf k}_\perp)^2}.
\end{equation}
and integrated them analytically over $dy$ and $d^2{\bf l}_\perp$. 
The remaining two integrations 
over $d^2{\bf k}_\perp$ have been done numerically with the 
built-in algorithms in Mathematica.
The result was stored as the ``IR-contribution''.
In Fig.\ref{diags01}, we present these amplitudes, corresponding to S0, S1 and S2. 
From Fig.\ref{diags01}, one can see how the IR-singularity subtraction works in practice.

In the reduced box-diagram (right in Fig.\ref{iniDia}), 
the struck parton momentum $k$
produces an UV-divergence
because this amplitudes falls off like $1/{\bf k}_\perp^2$ as $|{\bf k}_\perp| 
\rightarrow \infty$.
This results in a logarithmic UV-divergence.
Note that, since the original box diagram is UV-finite,
this UV-divergence is fictitious and is 
caused by the approximation for the effective vertex used herein, in which
the struck parton momentum was neglected and its 
propagator was replaced with ${1/Q^2}$, a procedure that  is
admissible as long as $k^2\ll Q^2$. We may handle this singularity by
using a cutoff $\Lambda^2$ of the order of $Q^2$. For $k^2\ge Q^2$,
the struck parton propagator should  behave as $1/{k^2}$. In principle, 
one may consider renormalization of this effective vertex to remove the dependence on 
$\Lambda$. However, we did not do this here because 
this term had only a weak dependence on the cutoff $\ln \Lambda^2\approx \ln Q^2$. 
In numerical calculations, we took this contribution in the form
\begin{equation}\label{LFGPD}
{\cal M}=\frac{(-\zeta) \, \overline{{\rm Tr} \,  \, [\dots]}/Q^2}
{\overline{{\rm Denominator}[\dots]}} \, 
h_{LF} h'_{LF} (y,{\bf l}_\perp) \times \frac{{\bf k}_\perp^2}
{({\bf k}_\perp^2+m^2)^2},
\end{equation}
and did the integration in ${\bf k}_\perp$ analytically: 
\be \label{UVfact}
B_\Lambda=\int\limits_{|{\bf k}_\perp | \leq \Lambda } d^2 {\bf k}_\perp  \, 
\frac{{\bf k}_\perp^2}{({\bf k}_\perp^2+m^2)^2}
= \pi  \left \{ \ln(1+\Lambda^2/m^2)- \frac{ \Lambda^2}{(m^2+\Lambda^2)}  \right \} .
\ee
The remaining expression was integrated over $dy d^2{\bf l}_\perp$ and stored as ``UV-contribution''.
Finally, after these parts were subtracted, the residual part of the 
amplitude was integrated numerically in
5 dimensions using MC method.

\begin{figure}
\centering
\epsfig{width=200pt,file=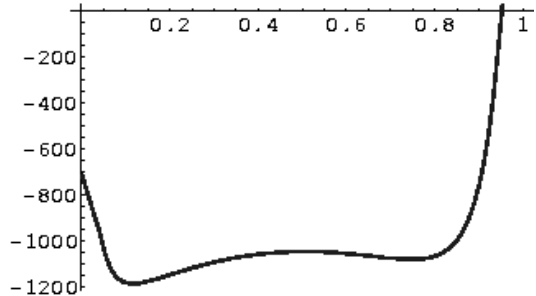}
\caption{Example of cumulative amplitude as
function of $\eta$  which
was integrated with 5D Monte-Carlo.}
\label{diags03}
\end{figure}

For the convenience of using Monte-Carlo, the integration over $y$ was
further reduced. Normally, the integration over $y$ would go over three regions,
e.g. $y<x<\zeta$, $x<y<\zeta$ and $x<\zeta<y$ if $x<\zeta$. These three 
integrals were rescaled, 
e.g. $y=x \eta$ in $y<x<\zeta$, $y=x+(\zeta - x) \eta$ in
$x<y<\zeta$ etc, so that $0<\eta<1$ and the three contributions, 
corresponding to these different regions, could be added together.
Their sum, as a function of $\eta$, was the cumulative amplitude to be 
actually integrated with 5D Monte-Carlo (see Fig.\ref{diags03}). 

The MC integration was organized in a series of bunches, each bunch
containing $N_i \sim 100 $ integrations using $M_i\sim 10,000$ points.
Each bunch is treated as a ``measurement" of the integral. When the test-points
are independent in the MC-integration, each measurement (bunch) is
independent from each other and the general principles of statistics can be 
applied.
For each bunch, thus, we estimated the dispersion $\sigma_i$ and the
mean $\mu_i$ of the measurement distribution. In this way,
we estimated the value of the integral as well as the statistical error
in each bunch from MC integrations.
All bunches finally were combined with appropriate weight factors to 
produce the final estimate of the integral and MC-integration error.
We tested our integration with different limits of
integration in $({\bf k}_\perp,{\bf l}_\perp)$ to see if
any accuracy is systematically lost due to finite bounds of integration in
${\bf k}_\perp$, ${\bf l}_\perp$. We observed that, except for rapid decrease of 
the MC-efficiency,
no noticeable change occurred when the integration region was enlarged.
Special attention was paid to $x=\zeta$ point since many of the amplitudes
become singular in this case. Although all of these singularities are removable,
we explicitly found the analytical limit in each expression for
$x\rightarrow \zeta$ either from the left or from the right, and accommodated
this in the final graphs as left (right) limit points at $x=\zeta$.

\begin{figure}
\begin{center}
\epsfig{width=200pt,file=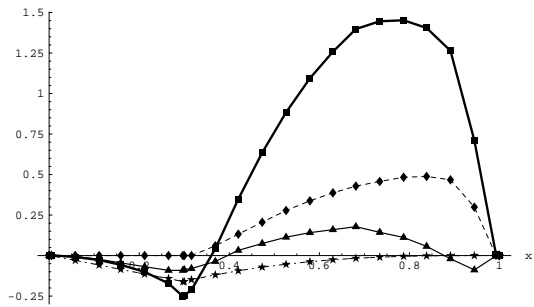}
\end{center}\vspace{0.3cm}
\caption{The plot of all contributions to scale 
for $\zeta=0.3$. The final result, ``IR-contribution",
``UV-contribution" and the 5D residual integral are denoted by
boxes, diamonds, stars and triangles, respectively. }
\label{final03}
\end{figure}

Thus, with our program we generated three different
numbers for each point $x,\zeta$. These were 
``IR-contribution", ``UV-contribution" and
the residual integrated amplitude. These were added to produce the final result 
\be
{\cal M}_{\rm final}={\cal M}_0+f({\cal M}_{IR} + {\cal M}_{res} + B_\Lambda {\cal M}_{UV}),
\ee
where ${\cal M}_0$ is the 0th-order amplitude, ${\cal M}_{IR}$ is IR-piece,
${\cal M}_{UV}$ is UV-piece of the amplitude and $B_\Lambda$ is the factor shown in 
Eq.~(\ref{UVfact}) with $\Lambda^2\approx Q^2$, where UV-cutoff 
$\Lambda \approx 5$\,GeV was typically used.
${\cal M}_{res}$ is the MC-integrated remaining contribution.
These results for $\zeta=0.3$ are plotted in Fig.\ref{final03}.

One final note is in order. In the LFBS approach of Ref.\cite{Choi:2001fc}, 
there is
a free parameter $G_\pi$ which can be adjusted to fit the pion form factor as 
\begin{equation}\label{sumX}
\int^1_{0} {dx} \, 
{\cal F}_{\pi}(\zeta, x, t)
= (1- \zeta/2) \, F_\pi(t) \, .
\end{equation}
In principle, $G_\pi$ may be allowed to depend on 
$t$ and $\zeta$. In that case one should solve for $G_\pi$ from
Eq.~(\ref{sumX}). To facilitate this process, we note that $G_\pi$ appears 
at most linearly from the non-valence final-state vertex in the final expressions. 
In our numerical procedure we explicitly separated $G_\pi^0$ and
$G_\pi^1$ parts of the amplitudes and computed separately
${\cal M}^0_f$ and  ${\cal M}^1_f$ so that
\be
{\cal M}_f={\cal M}^0_f+G_\pi {\cal M}^1_f.
\ee
Then $G_\pi$ can be easily found from Eq.~(\ref{sumX}) by solving a linear equation.
In our calculations, however, we did not solve exactly for $G_\pi(t,\zeta)$, 
but used $G_\pi = 0.32$ following the work of Ref.\cite{Choi:2002ic} where
the justification can be found. Similar to Ref.\cite{Choi:2002ic}, we
find that the condition (\ref{sumX}) is satisfied well for all values of
$\zeta$ even with a constant $G_\pi$.

\end{appendix}

\end{document}